\documentclass[fleqn]{article}
\usepackage{longtable}
\usepackage{graphicx}

\begin {document}
\begin{flushleft}
{\LARGE
{\bf Electron impact excitation rates for transitions in \\Mg V$^{\star}$}
}\\

\vspace{1.5 cm}

{\bf {Kanti  M.  ~Aggarwal and  Francis   P.   ~Keenan}}\\ 

\vspace*{1.0cm}

Astrophysics Research Centre, School of Mathematics and Physics, Queen's University Belfast, Belfast BT7 1NN, Northern Ireland, UK\\ 
\vspace*{0.5 cm} 

e-mail: K.Aggarwal@qub.ac.uk \\

\vspace*{0.20cm}

Received  15 July 2016.  Accepted 27 September  2016. \\

\vspace*{1.5cm}

PACS Ref: 31.25 Jf, 32.70 Cs,  34.80 Dp, 95.30 Ky

\vspace*{1.0 cm}

\hrule

\vspace{0.5 cm}
$\star$ Tables 2, 4, and 5 are available only in the electronic version as supplementary material.

\end{flushleft}

\clearpage


\begin{abstract}

Energy levels,  radiative rates (A-values) and lifetimes, calculated with the {\sc grasp} code, are reported  for an astrophysically important O-like ion Mg~V. Results are presented for transitions among the lowest 86 levels belonging to the 2s$^2$2p$^4$, 2s2p$^5$, 2p$^6$, and 2s$^2$2p$^3$3$\ell$ configurations.  There is  satisfactory agreement with earlier data for most levels/transitions, but scope remains for improvement.   Collision strengths are also calculated, with the {\sc darc} code, and the results obtained are comparable for most transitions (at energies above thresholds) with earlier work using the DW code.  In thresholds region, resonances have been resolved in a fine energy mesh to determine values of effective collision strengths ($\Upsilon$) as accurately as possible. Results are reported for all transitions at temperatures up to 10$^6$~K, which should be sufficient for most astrophysical applications. However, a comparison with earlier data reveals  discrepancies of up to two orders of magnitude for over 60\% of transitions, at all temperatures. The reasons for these discrepancies are discussed in detail.

\end{abstract}

\clearpage

\section{Introduction}

Spectral lines of Mg~V are frequently observed from a wide range of  astrophysical plasmas, including the Sun (see for example, \cite{fd}, \cite{sand},  \cite{tom}, \cite{dhb}),  other stars \cite{pens}, and planetary nebulae \cite{rus}. Similarly, Mg~V lines have also been detected from laboratory plasmas (\cite{hg}, \cite{kart}). Since many emission  line pairs of Mg~V are density or temperature sensitive \cite{pry}, they provide excellent diagnostics for astrophysical plasmas. However, the analysis of observations requires information on atomic data, such as  energy levels,  radiative rates (A-values), and excitation rates which are obtained from the collision strengths ($\Omega$). Measurements of energy levels for Mg~V have been compiled and assessed by \cite{mz} and their recommended values are available on the  NIST (National Institute of Standards and Technology)  website {\tt http://www.nist.gov/pml/data/asd.cfm}. However, to our knowledge there are no measurements for the  A-values and $\Omega$, and therefore theoretical results for these are required.

Early calculations for $\Omega$ (and effective collision strengths $\Upsilon$) for Mg~V were performed by \cite{cmz} and \cite{bz}, but only for transitions among levels of the $n$=2 configurations. Their results are not sufficient for the analysis of observed lines in the x-ray region, because these arise from levels of the $n$=3 configurations \cite{ble}. Therefore, \cite{hud} extended the work to include 37 levels of the 2s$^2$2p$^4$, 2s2p$^5$, 2p$^6$, 2s$^2$2p$^3$3s, and 2s$^2$2p$^3$3p configurations. They adopted the Breit-Pauli $R$-matrix method of \cite{scot}, although their calculations were primarily  in $LS$  coupling (Russell-Saunders or spin-orbit coupling). However, for a lowly ionised ion, such as Mg~V, this method should not affect the accuracy of the calculated data, because relativistic effects are not too important. Furthermore, \cite{hud} resolved resonances in a narrow energy mesh and presented values of $\Upsilon$ over a wide temperature range, up to 10$^7$~K. However, a much larger calculation involving 86 levels of the  2s$^2$2p$^4$, 2s2p$^5$, 2p$^6$, and 2s$^2$2p$^3$3$\ell$ configurations, has already been reported by \cite{ble}. For the calculation of energy levels they adopted the {\em SuperStructure} (SS) code and considered up to 24 configurations ($n \le$ 4). However, for the scattering calculations, their wavefunctions were rather basic (among the above listed six configurations) and hence the discrepancies between  the calculated energies and the NIST compilation are significant for several levels, in both magnitude and orderings -- see table 1 of \cite{ble}, and note particularly the position of the 2p$^6$~$^1$S$_0$ level. More importantly, they calculated their data for $\Omega$ at only five energies, in the 10--50~Ryd range, with the {\em distorted wave} (DW)  method, and did not consider resonances in the thresholds region. Since resonances for transitions in Mg~V are very important (see section 5), the results obtained for $\Upsilon$ from their calculated values of $\Omega$ are likely to be highly underestimated for many transitions, particularly  non-dipole allowed ones. 

Therefore, recently \cite{sst} have performed yet another calculation for  Mg~V. They have considered the same 86 levels of the six configurations as by \cite{ble}, but have made several improvements over their work. For the determination of atomic structure (i.e. to calculate energy levels and A-values) they adopted the multi-configuration Hartree-Fock (MCHF) code of \cite{mchf}, used non-orthogonal orbitals up to $n$=5 ($\ell \le$ 4), and included very large {\em configuration interaction} (CI), up to 5179. As a result, the accuracy of their calculated energy levels is much better -- see their table 1. Advantages of this methodology (over the use of orthogonal orbitals) include the determination of accurate energy levels, but avoidance of pseudo resonances in the subsequent scattering calculations, as faced by \cite{hud}. However, the inclusion of such a large CI is not feasible in the scattering calculations, due to computational limitations. Therefore, \cite{sst} had to make a compromise by reducing the CI to 427 configurations, and the energies obtained with this reduced model are not too different from the larger one -- see their table~1.  

For the calculation of $\Omega$, \cite{sst} adopted the B-spline $R$-matrix (BSR) code developed by \cite{zat}. They included a moderately large range of partial waves with angular momentum $J$ up to 29.5, but calculated $\Omega$ over a wide energy range up to 50~Ryd. To resolve resonances in the thresholds region they considered up to as narrow an energy  mesh as 0.000~05~Ryd, and reported values of $\Upsilon$ up to T$_e$ = 10$^6$~K. Therefore, their calculated results not only cover a larger range of transitions, but  would be expected to be the most accurate. Hence, there should be no strong  reason to perform yet another calculation for Mg~V. However, they conclude  that they observed a good agreement for most transitions between their results of $\Upsilon$ and those of \cite{hud}. We do not find this to be true (as discussed below), and hence have undertaken a further calculation.

In Fig. 1 (a, b and c) we show comparisons of $\Upsilon$ calculated by \cite{hud} and \cite{sst}, referred to as RM1 and RM2 for convenience, at three temperatures of 10$^4$, 10$^5$, and 10$^6$~K. These are shown as the ratio (R) of $\Upsilon_{RM2}$/$\Upsilon_{RM1}$, with negative values indicating $\Upsilon_{RM1}$/$\Upsilon_{RM2}$. Two conclusions can be easily drawn from this figure. First,  the discrepancies are large (up to two orders of magnitude) for a significant number of transitions ($\sim$80\%), and second the agreement worsens with increasing T$_e$. It is interesting to note that at T$_e$ = 10$^6$~K, the $\Upsilon_{RM1}$ of \cite{hud} are much larger than those of \cite{sst}, whereas the opposite would be expected,  because the RM2 calculations included more levels, and hence significantly more resonances. We discuss these comparisons in more detail below to try to understand the reasons for the discrepancies. 

At T$_e$=10$^4$~K, the $\Upsilon_{RM2}$ are larger than $\Upsilon_{RM1}$ for many transitions, by up to two orders of magnitude. Differences between the two sets of $\Upsilon$ are greater than 20\% for 83\% of transitions. For only a few,  $\Upsilon_{RM1}$ are larger than $\Upsilon_{RM2}$, by a maximum of a factor of 40. Although 10$^4$~K is a comparatively  low temperature at which the positions and magnitudes of resonances  are very important, the differences between the two sets of $\Upsilon$ are still very large, considering that both calculations resolved resonances in a fine energy mesh, i.e. 0.0008~Ryd in RM1 and  0.000~05~Ryd in RM2. The most noticeable discrepancies seen in Fig. 1a are for the transitions: 5--10 (2s$^2$2p$^4$~$^1$S$_0$ -- 2s$^2$2p$^3$3s~$^5$S$^o_2$ : R = 430, out of scale), 5--31 (2s$^2$2p$^4$~$^1$S$_0$ -- 2s$^2$2p$^3$3p~$^3$F$_2$ : R = 60), 5--32 (2s$^2$2p$^4$~$^1$S$_0$ -- 2s$^2$2p$^3$3p~$^3$F$_3$ : R = 92), 5--33 (2s$^2$2p$^4$~$^1$S$_0$ -- 2s$^2$2p$^3$3p~$^3$F$_4$ : R = 126), and 5--34 (2s$^2$2p$^4$~$^1$S$_0$ -- 2s$^2$2p$^3$3p~$^1$F$_3$ : R = 60). All these  are forbidden transitions and therefore the number and magnitude of resonances may considerably affect the determination of $\Upsilon$. We also note that the ordering (labelling) of these levels is that given in Table~1 (but slightly different from RM1 and RM2, which position the 2s$^2$2p$^3$3s~$^5$S$^o_2$ level at number 11), but  there is no ambiguity in comparing results. Since the discrepancy is greatest for the 5--10 transition, we focus our attention on this alone. The background value of $\Omega$ ($\Omega_B$) for this transition in the thresholds region is $\sim$10$^{-3}$ in our calculations (see section 4)  and is $\sim$10$^{-8}$ at energies well above the thresholds, comparable to the DW results of \cite{ble}, and resonances are neither too numerous nor too large in magnitude. Subsequently, our values of $\Upsilon$ for this transition are  $\sim$10$^{-3}$ (see  Table~5 in section 5). However, the corresponding $\Upsilon$ of RM1 and RM2 are lower and higher, respectively, than our result by about an order of magnitude, and hence the discrepancies. If \cite{sst} have some very large resonances close to the threshold, then their $\Upsilon$ may be considerably higher towards the lower end of T$_e$. Indeed this is the case as may be noted from Fig. 1 (b and c), because the discrepancies have decreased with increasing temperature.  

At T$_e$=10$^5$~K, the RM1 and RM2 values of $\Upsilon$ differ for 79\% of transitions by over 20\%, and in most cases the results of \cite{sst} are larger, as seen in Fig. 1b. However, at T$_e$=10$^6$~K the trend has reversed,  as for most transitions the $\Upsilon$ of \cite{hud} are larger by up to three orders of magnitude (note that a few transitions are out of scale in Fig.~1c). This is not expected (as noted already) and the greatest discrepancies are for the transitions: 16--24 (2s$^2$2p$^3$3p~$^5$P$_1$ -- 2s$^2$2p$^3$3s~$^3$P$^o_1$ : R = 980), 16--25 (2s$^2$2p$^3$3p~$^5$P$_1$ -- 2s$^2$2p$^3$3s~$^3$P$^o_2$ : R = 980), and 18--34 (2s$^2$2p$^3$3p~$^5$P$_3$ -- 2s$^2$2p$^3$3p~$^1$F$_3$ : R = 820).  The first two are inter-combination transitions (but very weak as their f-values are 3.6$\times$10$^{-8}$ and 1.6$\times$10$^{-7}$, respectively) whereas the third is  forbidden. A temperature of 10$^6$~K corresponds to $\sim$6.3~Ryd and the above  transitions are close to the highest level (37) considered by \cite{hud}. Hence, the contribution of resonances (if any) should not be too significant. Clearly, the $\Upsilon$ of \cite{hud} are  overestimated at high values of T$_e$,  because they calculated data for $\Omega$  only up to 28~Ryd, but determined $\Upsilon$  up to T$_e$ = 10$^7$~K, i.e. $\sim$63~Ryd. There is no mention in their paper if they extrapolated values of $\Omega$ to high energies, but one of the authors (Cathy Ramsbottom) confirms that they did not. An inspection of the  $\Upsilon$ values of \cite{hud} over a wide temperature range also indicates they are overestimated at high T$_e$, because for many transitions (such as: 16--24, 16--25, 17--34, 18--32, and 18--34) their results increase by (almost) up to  three orders of magnitude between the lowest (10$^3$~K) and the greatest (10$^7$~K) temperature. Values of $\Upsilon$ are not (generally) known to vary so largely. Additionally, these transitions are either weak inter-combination or forbidden,  and therefore their values of $\Omega$ and subsequently $\Upsilon$ should decrease  considerably with increasing energy/temperature. This is indeed the case with the results of  \cite{ble} and \cite{sst}, or in our calculations -- see Table~5 in section 5. In conclusion, we may confidently state that the  results of \cite{hud} are  overestimated, at least at high T$_e$. We therefore do not discuss their results further.

The temperature range below 10$^6$~K is most important for the analysis of astrophysical observations of Mg~V lines \cite{ble}, and unfortunately there are large discrepancies between the values of $\Upsilon$ calculated by  \cite{hud} and \cite{sst}, as shown in Fig. 1 and discussed above. Additionally, we have reason to suspect the accuracy of the  results for $\Upsilon$ reported by \cite{sst}, but will  discuss these in detail  more appropriately  in section 5. Therefore, we have performed one more calculation for this important ion, and for consistency as well as for practical (computational) reasons include the same 86 levels as by \cite{sst}. However, our approach is different (but similar) from their work because for the generation of wavefunctions  we have adopted the  fully relativistic {\sc grasp} (General-purpose Relativistic Atomic Structure  Package) code,  originally  developed by  \cite{grasp0}, but  significantly revised by Dr. P. H. Norrington.   It is a fully relativistic code,  based on the $jj$ coupling scheme. Further higher order relativistic terms arising from the Breit interaction and QED (quantum electrodynamics) effects (vacuum polarisation and Lamb shift)  have also been included in the same way as  described in the original version. Additionally, we have used the option of {\em extended average level} (EAL),  in which a weighted (proportional to 2$j$+1) trace of the Hamiltonian matrix is minimised. This produces a compromise set of orbitals describing closely lying states with moderate accuracy, and generally yields results comparable to other options, such as {\em average level} (AL).   Finally, energy levels and radiative data determined with {\sc grasp} are generally comparable with those obtained with other codes, such as CIV3, FAC, MCHF, and SS, provided similar CI is included.

A particularly useful feature of the adopted version is that it also has provisions for listing the $LSJ$ designations of the levels/configurations, apart from the usual $jj$ nomenclature of the relativistic codes. This feature helps in correctly identifying the states/levels and facilitates direct comparisons with other calculations. Another useful advantage of this version is that its output can be directly linked to the collisional code (i.e. the {\em Dirac atomic R-matrix code}, DARC), adopted here  for the scattering calculations. This code also  includes the relativistic effects in a systematic way, in both the target description and the scattering model, because it is based on the $jj$ coupling scheme, and uses the  Dirac-Coulomb Hamiltonian in the $R$-matrix approach. However, the code does not include the Breit and QED corrections, and hence the target energies obtained are slightly different (and comparatively less accurate) than from {\sc grasp}.  A disadvantage  of this code is that the calculations are more time consuming because the size of the Hamiltonian matrix is large, due to the inclusion of fine-structure in the definition of channel coupling. On the other hand, an advantage of the code is that resonances arising in between the fine-structure levels of a state can be included, and hence improving the further calculations of effective collision strengths -- note particularly the ground state  levels (2s$^2$2p$^4$~$^3$P$_{0,1,2}$) in Table~1.

 Both the atomic structure and scattering codes, i.e. {\sc grasp} and {\sc darc}, are hosted at the website: \\{\tt http://amdpp.phys.strath.ac.uk/UK\_APAP/codes.html}.  Finally, we stress that although we have adopted fully relativistic codes for calculations of both atomic structure and collisional parameters, the relativistic effects are not too important for a comparatively light ion such as  Mg~V.


\section{Energy levels}

Determining the atomic structure of Mg~V is not  simple and straightforward, as also discussed by \cite{ble} -- see their table~1. For the same reason, \cite{sst} considered a large CI and adopted the route of pseudo (correlation) and non-orthogonal orbitals. Our DARC code does not allow such a provision and therefore, as in most of our work on many other ions, we use the orthogonal orbitals. However, we consider 12 configurations, namely  2s$^2$2p$^4$, 2s2p$^5$, 2p$^6$, 2s$^2$2p$^3$3$\ell$, 2s2p$^4$3$\ell$, and 2p$^5$3$\ell$, which generate 226 levels in total. Inclusion of these configurations considerably improves the accuracy of the energy levels and is a good `compromise' for the subsequent calculations of scattering parameters. Furthermore,  as stated above, the calculations for energy levels have been performed in an `extended average level' (EAL) approximation. The contribution of Breit and QED (quantum electrodynamic) effects has also been included,  but has been found to be insignificant, as expected.

In Table~1 we compare our GRASP energies with the measurements compiled by NIST, and with other calculations, namely MCHF1 \cite{sst}, MCHF2 \cite{tff}, and SS \cite{ble}.  It may be noted that the MCHF2 energies are listed from the website  {\tt  http://nlte.nist.gov/MCHF/view.html}, and are {\em slightly} different from those given in \cite{tff}. As expected, the MCHF1 and MCHF2 energies are comparable, in both magnitude and orderings, and the slight differences observed for a few levels (such as 23--25) are due to the `reduced' CI model considered by \cite{sst} for the collisional work. Otherwise, the discrepancies between the two sets of energies are (even) smaller, as may be seen in table~1 of  \cite{sst}.  The SS energies calculated by \cite{ble} are in worse agreement with NIST, in both magnitude and orderings, because of the limited CI, and this has already been commented on by these authors. Our calculated energies with the {\sc grasp} code are comparatively more accurate than those of SS, because of the larger CI included. However, there is scope for improvement, but keeping in  mind our further (and more important) calculations for $\Omega$, it is a reasonable compromise as already stated. Differences with the NIST compilation are within 0.1~Ryd for most levels, except 12 (2p$^6$~$^1$S$_0$) which differs by 0.4~Ryd and its position is also different. A few more levels for which the orderings (slightly) differ between our work and NIST are 13/14, 65--67, and 69--71, but all these  are closely lying, i.e. $\Delta$E is very small. Finally, we note that the 2p$^3$3d~$^1$S$^o_0$ and 2p$^3$3p~$^1$S$_0$ levels (58 and 60) have reverse orderings between GRASP and MCHF1 (and MCHF2). However, a closer look at the $\Upsilon$ results of \cite{sst} indicates that their ordering for these two levels is the same as ours, and this has been based on the trend of the allowed and forbidden transitions related to these levels.

 Some of the  levels listed in Table~1  (under the GRASP column) are not strictly in the order of increasing energy -- see for example, levels 13--15, 35/36, 38/39, 61--63, and 65/66. This is because the contributions  of Breit and QED effects (included in {\sc grasp} but not in {\sc darc}), although not very significant as already stated, alters the orderings of the Coulomb energies, used in the calculations of $\Omega$ and subsequently $\Upsilon$. Therefore, the orderings of levels for all atomic parameters, namely energies, A-values, $\tau$, $\Omega$, and $\Upsilon$, are consistent with the one listed in Table~1, although some anomalies for energies are noticeable.

\section{Radiative rates and lifetimes}

Besides energy levels, we have also calculated  A-values  for four types of transitions, namely electric dipole (E1), electric quadrupole (E2), magnetic dipole (M1), and  magnetic quadrupole (M2), because these are required for  plasma modelling as well as for the determination of lifetimes ($\tau$), a measurable quantity. Our calculated energies/wavelengths ($\lambda$, in $\rm \AA$), radiative rates (A$_{ji}$, in s$^{-1}$), oscillator strengths (f$_{ij}$, dimensionless), and line strengths (S, in atomic unit = 
 6.460$\times$10$^{-36}$ cm$^2$ esu$^2$) are listed in Table~2 for all  E1 transitions, but only among the lowest 86 levels of  Mg~V, although calculations have been performed for 226 levels (see section 2). This is because our subsequent calculations for $\Omega$ and $\Upsilon$ are only for these 86 levels.  The {\em indices} used  to represent the lower and upper levels of a transition are defined in Table~1. Furthermore, for the E2, M1, and M2  transitions  only the A-values are listed in Table~2, because corresponding data for f-values can be easily obtained through the following equation, i.e.  

\begin{equation}
f_{ij} = \frac{mc}{8{\pi}^2{e^2}}{\lambda^2_{ji}} \frac{{\omega}_j}{{\omega}_i} A_{ji}
 = 1.49 \times 10^{-16} \lambda^2_{ji}  \frac{{\omega}_j}{{\omega}_i}  A_{ji} 
\end{equation}
where $m$ and $e$ are the electron mass and charge, respectively, $c$ the velocity of light,  and $\omega_i$ and $\omega_j$  the statistical weights of the lower ($i$) and upper ($j$) levels, respectively. Finally,  results are only listed in the length  form, which are (generally) considered to be comparatively more accurate. However, ratio of the velocity and length forms of A-values for E1 transitions are also listed in this table as a last column, as it gives an indication of the accuracy of the calculated data.

Similar results, particularly for the E1 transitions, are available in the literature, mainly by \cite{ble} and \cite{sst}, and the latter have also shown comparisons with earlier existing results. Since the A-values of \cite{sst} are comparatively more accurate (because of the inclusion of larger CI and hence more accurate determination of energy levels and subsequent other atomic parameters) we compare our results with them (MCHF1) alone. In Table~3 are compared the f-values for transitions among the lowest 13 levels, and as expected, the agreement between our GRASP and their MCHF1 results is (generally) within 20\%, particularly for those which have significant magnitude. However, for a few weaker transitions, such as 4--7, 4--13, and  5--13, discrepancies are up to a factor of two, whereas for  12--13 (f$\sim$ 10$^{-8}$) the two sets of f-values differ by three orders of magnitude. Such discrepancies for weak(er) transitions, between any two calculations differing in CI and/or methodology, are quite common. Similarly, the ratio (R) of the velocity and length forms are generally within 20\% of unity, but are larger for the weak(er) transitions, as also shown in fig.1 of \cite{sst}. Finally, we  stress that the f- (A-) values of \cite{sst} may be comparatively more accurate (for the reasons already stated), but our results are given again in Table~2 for two reasons, i.e. firstly for consistency with our subsequent results of $\Upsilon$ (see section 5) and secondly, for the addition of the E2, M1, and M2 transitions, which were not considered by \cite{sst}. 

Inclusion of A-values for all types of transitions results in a more accurate determination of the lifetime ($\tau$ = 1.0/${\sum_{i}} A_{ji}$). This is a measurable quantity and helps in assessing the accuracy of the calculated data. Unfortunately, no measurements of $\tau$ are available for the levels of Mg~V, but the theoretical MCHF2 results of  \cite{tff} are available on the website {\tt  http://nlte.nist.gov/MCHF/view.html}. Therefore, in Table~1 we have also listed our and their calculated values of $\tau$ for all levels. For most, the two sets of $\tau$ agree within $\sim$20\%, but discrepancies are larger (by up to a factor of two) for a few levels, such as 40--45 for which the GRASP results are lower. These differences are a direct consequence of the corresponding discrepancies  in the A-values. For example, for level 40 (2p$^3$3p~$^3$S$_1$) the dominant A-value is for the 6--40 E1 transition, for which our result is 2.67$\times$10$^9$ (s$^{-1}$) whereas that from MCHF2 is 6.55$\times$10$^8$ (s$^{-1}$). This is a weak transition (f $\sim$ 10$^{-3}$) for which the differences in f- and A-values have already been noted. The same analysis applies to other levels where the discrepancies are large. In the absence of measurements or other theoretical results, it is difficult to say with confidence which set of $\tau$ values is more accurate.

\section{Collision strengths}
As stated in section ~1, we have adopted  the  relativistic  {\sc darc} code for calculating  $\Omega$,  a symmetric and dimensionless quantity. The $R$-matrix radius adopted for  Mg~V is 8.0 atomic units, and 62  continuum orbitals (NRANG2) have been included for each channel angular momentum in the expansion of the wavefunction. This large expansion allows us  to compute results for $\Omega$ up to an energy of  30 Ryd. Furthermore,  the maximum number of channels generated for a partial wave is 388, but considering the high value of NRANG2  the size of the (largest) Hamiltonian (H) matrix becomes 24~079. For calculating $\Omega$ we have considered all partial waves with angular momentum $J \le$ 40.5, sufficient for  convergence  for a majority of transitions and at all energies.  However, to improve the accuracy further we have included the contribution of higher neglected partial waves through the Coulomb-Bethe  \cite{ab} and  geometric series  approximations for allowed and forbidden transitions, respectively. 

To our knowledge no measurements for $\Omega$ have been made for the transitions of Mg~V, and the only theoretical results available for comparison  with  our data are those of \cite{ble}. Some differences with their results are expected, because they have adopted the DW method and (more importantly) their wavefunctions are basic, as already stated in section 1. Nevertheless, in Fig.~2a we show the comparisons of $\Omega$ for three forbidden transitions, namely 1--2 (2s$^2$2p$^4$~$^3$P$_2$ -- 2s$^2$2p$^4$~$^3$P$_1$), 1--22 (2s$^2$2p$^4$~$^3$P$_2$ -- 2s$^2$2p$^3$($^4$S)3p~$^3$P$_2$), and 1--42 (2s$^2$2p$^4$~$^3$P$_2$ -- 2s$^2$2p$^3$($^2$D)3p~$^3$P$_2$). These transitions are selected because of their comparatively larger magnitudes. For the 1--2 and 1--22 transitions there are no discrepancies between the two sets of $\Omega$, but for 1--42  our results are lower by up to 15\%. We stress  that for this (and most other) forbidden transitions, the values of $\Omega$ have fully converged within our partial waves range. In fact, for this transition,  $\Omega$ has converged within the lowest 20 partial waves alone, and therefore the discrepancies noted here are because of the differences in methodology and the wavefunctions. Similar comparisons are shown in Fig.~2b for three allowed transitions, namely 1--6 (2s$^2$2p$^4$~$^3$P$_2$ -- 2s2p$^5$~$^3$P$^o_2$), 1--7 (2s$^2$2p$^4$~$^3$P$_2$ -- 2s2p$^5$~$^3$P$^o_1$), and 4--9 (2s$^2$2p$^4$~$^1$D$_2$ -- 2s2p$^5$~$^1$P$^o_1$). For these (and many other) allowed transitions there are no discrepancies between our results and those of \cite{ble}, because the f-values are also comparable. Therefore, we have confidence in our calculations for $\Omega$ and in Table~4 we list our results for  all  transitions from the ground level over the 9--30~Ryd energy range. These data should be useful for future comparisons.

\section{Effective collision strengths}

Our listed values of $\Omega$ in Table~4 are not very useful for applications. This is because, in most plasmas, electrons do not have mono-energetic velocities but rather have a distribution, most often a {\em Maxwellian} one. Furthermore, in the thresholds region values of $\Omega$ do not vary as smoothly as shown in Fig.~2. For most transitions,  particularly the non-dipole allowed ones, the thresholds region is full of numerous closed-channel (Feshbach) resonances, as shown by \cite{hud} in their fig. 3 for three transitions, and by \cite{sst} in their figs. 3 and 4 for two transitions. However, the importance of resonances  cannot be fully appreciated from their figures, because they  cover a very wide energy range, up to 50~Ryd, whereas the thresholds region is below 9~Ryd.  Therefore, in Fig. 3 (a, b and c) we show resonances for three transitions, namely 1--2 (2s$^2$2p$^4$~$^3$P$_2$ -- 2s$^2$2p$^4$~$^3$P$_1$), 1--3 (2s$^2$2p$^4$~$^3$P$_2$ -- 2s$^2$2p$^4$~$^3$P$_0$), and 4--9 (2s$^2$2p$^4$~$^1$D$_2$ -- 2s2p$^5$~$^1$P$^o_1$). The first two are forbidden whereas the third is  allowed. It is clear from these figures that resonances are not only dense but also have large magnitudes. For this reason,  \cite{sst} resolved these in a fine energy mesh, up to  0.000~05~Ryd in some energy regions. In most of the thresholds region, our energy mesh is  0.0001~Ryd and in total we have calculated $\Omega$ at 47~114 energies, compared to \cite{sst} who used 34~352 energy points. 

From the data for $\Omega$,  the effective collision strengths $\Upsilon$ are determined as follows: 

\begin{equation}
\Upsilon(T_e) = \int_{0}^{\infty} {\Omega}(E) \, {\rm exp}(-E_j/kT_e) \,d(E_j/{kT_e}),
\end{equation}
where $k$ is Boltzmann constant, T$_e$  the electron temperature in K, and E$_j$  the electron energy with respect to the final (excited) state. This value of $\Upsilon$ is
related to the excitation q(i,j) and de-excitation q(j,i) rates as follows:

\begin{equation}
q(i,j) = \frac{8.63 \times 10^{-6}}{{\omega_i}{T_e^{1/2}}} \Upsilon \, {\rm exp}(-E_{ij}/{kT_e}) \hspace*{1.0 cm}{\rm cm^3s^{-1}}
\end{equation}
and
\begin{equation}
q(j,i) = \frac{8.63 \times 10^{-6}}{{\omega_j}{T_e^{1/2}}} \Upsilon \hspace*{1.0 cm}{\rm cm^3 s^{-1}},
\end{equation}
where $\omega_i$ and $\omega_j$ are the statistical weights of the initial ($i$) and final ($j$) states, respectively, and E$_{ij}$ is the transition energy. Results for these rates are required in the modelling of plasmas.

Our calculated values of  $\Upsilon$ are listed in Table~5 up to T$_e$ = 10$^{6}$~K, well above  the temperature of maximum abundance in ionisation equilibrium for Mg~V, i.e. 10$^{5.5}$~K  \cite{pb}. As  discussed in section~1, the most recent, extensive and perhaps best available corresponding data for $\Upsilon$ are those of \cite{sst}.  Therefore, we undertake a detailed  comparison  with their results.

In Fig. 4 (a, b and c) we compare our results ($ \Upsilon_{DARC}$) with those of \cite{sst}  ($\Upsilon_{TS}$) in the form of the ratio R = $\Upsilon_{TS}$/$\Upsilon_{DARC}$, with negative values of R representing $\Upsilon_{DARC}$/$ \Upsilon_{TS}$, i.e.  $\Upsilon_{DARC}  >  \Upsilon_{TS}$. These comparisons of $\Upsilon$ are for all transitions among the lowest 86 levels, and   at  three temperatures of 10$^4$, 10$^5$, and 10$^6$~K. Furthermore, they are similar to those shown in Fig.~1, but the lower levels (I) have been replaced by the upper ones (J), as this provides a clearer picture of the similarities or differences among transitions  up to level(s) J. 

From Fig.~4 two conclusions can be clearly drawn, i.e. the discrepancies between the two sets of $\Upsilon$ worsen with increasing temperature, and for a majority of transitions $\Upsilon_{DARC}  >  \Upsilon_{TS}$, by up to two orders of magnitude. At 10$^4$~K, there is reasonable agreement between the two sets of $\Upsilon$ for transitions among the lowest $\sim$25 levels alone, but at 10$^5$ and 10$^6$~K, this reduces to J = $\sim$20 and $\sim$15, respectively. Since the \cite{sst} and our calculations are based on the $R$-matrix method, resolve resonances in a fine energy mesh, and have taken care to obtain converged values of $\Omega$ at all energies and for all (type of) transitions, a much better agreement between the $\Upsilon$ values is expected. Unfortunately, the discrepancies are very large and therefore we discuss the potential reasons in detail.

As most of the discrepancies between the two sets of $\Upsilon$ are for transitions among the higher levels of Mg~V, in Table~6 we compare our  results with those of \cite{sst} at three  temperatures of 10$^4$, 10$^5$, and 10$^6$~K, for all transitions with upper level 86. This comparison alone should, we believe, be sufficient to obtain a clear(er) idea about the discrepancies, and for the same reason we have also listed the f-values (for allowed)  and $\Omega$ values for all transitions at three energies of 9, 12, and 15~Ryd. The energy of 9~Ryd is close to that of  threshold  for level 86 (i.e. $\sim$8.3~Ryd -- see Table~1) and  resonances play no part in determining the $\Upsilon$ for these transitions. Hence, the values of $\Omega$ and $\Upsilon$ should increase/decrease with increasing energy/temperature, depending on the (type of) transition. Indeed, this is the case in our calculations as may be seen from Table~6. However, the $\Upsilon$ of \cite{sst}  show a ``hump" at 10$^5$~K for almost all transitions, i.e. between 10$^4$ and 10$^6$~K they first increase (by up to a factor of two) and then decrease. This trend in their $\Upsilon$ results, except for the strong allowed transitions (such as 5--86, 48--86, and 60--86), indicates the presence of pseudo resonances in their calculated values of $\Omega$. To specifically avoid this problem they adopted the non-orthogonal orbitals in the construction of wavefunctions and employed the BSR code for the calculations of $\Omega$, but unfortunately their results are not correct. We discuss this further below.  

Apart from pseudo resonances which affect the values of $\Upsilon$ at higher temperatures, the \cite{sst} results appear to be highly underestimated  at low(er) temperatures. This is  (again) indicated by in particular the above-mentioned strong allowed transitions -- see table~4 of \cite{sst} -- because between 10$^{3.2}$ and 10$^6$~K (i.e. within 6.3~Ryd) their results {\em increase} by up to two orders of magnitude for several transitions, such as 4/5/9/18/26/51/55/56 to 86, which include both allowed  (4/5/55 to 86) and forbidden  (9/18/26/51/56 to 86). Such a large increase in $\Upsilon$ values is (generally) not found, particularly for transitions which do not have any resonance contributions. We do not observe this in our calculations for Mg~V and nor have, for example, \cite{mfr} in their work on S~III, which is also a light, lowly-ionised ion. The only transitions for which our values of $\Upsilon$ increase by up to a factor of 100 over this temperature range are three forbidden (12--72, 18--26, and 26--38) and two inter-combination, namely 12--71 (f = 3.5$\times$10$^{-7}$) and 12--73 (f = 9.5$\times$10$^{-10}$), and the reason for the increase is understandable, i.e. due to the large number of resonances.  Furthermore, for the 5/48/60--86 allowed transitions, the f-values in the two calculations are comparable for 48--86, but differ by $\sim$50\% for the other two. However, the $\Upsilon$ of \cite{sst} are comparable with our results at 10$^5$ and 10$^6$~K, but are significantly underestimated at 10$^4$~K (by up to a factor of three) as shown in Table~6.

For some forbidden transitions listed in Table~6, the $\Upsilon$  of \cite{sst} increase (by up to two orders of magnitude --  see their table~4 for full results) with increasing temperature whereas our values decrease (by a maximum factor of three), and are in accordance with the corresponding behaviour of $\Omega$. Examples of such transitions are: 9/17/34/37/38/68/74 to 86. A  part of the reason for these anomalies may be that \cite{sst} calculated values of $\Omega$ up to an energy of 50~Ryd, but the partial waves range included was only up to 29.5, although they did include the contribution of higher neglected partial waves with top-ups in the same way as in our calculations. In comparison we included all partial waves up to $J$ = 40.5 to calculate $\Omega$ up to 30~Ryd,  sufficient to determine $\Upsilon$ up to 10$^6$~K, as already discussed. Some transitions, such as 34--86 (2p$^3$3p~$^1$F$_3$ -- 2p$^3$3d~$^1$P$^o_1$), 68--86 (2p$^3$3d~$^1$P$^o_1$ -- 2p$^3$3d~$^1$P$^o_1$), and 74--86 (2p$^3$3d~$^1$F$^o_3$ -- 2p$^3$3d~$^1$P$^o_1$) converge very slowly, particularly towards the higher end of the energy range. We have undertaken test calculations with top-ups at J = 20.5, 30.5, and 40.5 to assess what differences they make. Values of $\Omega$ obtained with top-up at 20.5 are larger by up to 50\% than at 30.5, for some transitions (such as those mentioned here) at an energy of 30~Ryd. However, a similar comparison with top-ups at 30.5 and 40.5 shows a maximum difference of $\sim$10\%. However, similar comparisons at 50~Ryd will almost certainly be different and most likely the values of $\Omega$ in the calculations of \cite{sst} are overestimated, although this cannot be confirmed as their data are not available. Nevertheless, it does explain  (to some extent) why for some forbidden transitions their $\Upsilon$ results  increase with increasing temperature. Therefore, for many transitions their $\Upsilon$ values appear to be underestimated towards the lower end of the temperature range but overestimated at the higher end, and hence the discrepancies shown in Fig.~4 and Table~6. However, \,{\em if}\, the $\Upsilon$ of \cite{sst} are overestimated, particularly at higher temperatures,  because of the presence of pseudo resonances and/or overestimation in their calculated values of $\Omega$ (as discussed earlier), then why are our results still larger for many transitions, as shown in Fig.~4. To understand this anomaly we need to undertake some additional comparisons.

In Fig.~5 (a, b and c) we compare the two sets of $ \Upsilon$ again, as in Fig.~4, but this time replacing the upper levels (J) with lower ones (I). It is clear from this figure that the maximum discrepancies are for transitions from lower levels below 12, and our results are invariably higher (by up to two orders of magnitude) at all temperatures. There are a few more transitions (such as 9 -- 52/53/54 and 23 --43) which are out of scale, but among those shown in Fig.~5 are two -- 12--28 (2p$^6$~$^1$S$_0$ -- 2p$^3$3p~$^3$D$_2$) and 12 -- 30 (2p$^6$~$^1$S$_0$ -- 2p$^3$3p~$^3$D$_3$) --  both forbidden,  for which our $\Upsilon$ are higher by a factor of 140 at 10$^6$~K. In Fig.~6 we show our $\Omega$ for the 12--30 transition alone. Although there are some resonances for this (and other similar) transitions, they are neither too dense nor too large in magnitude, and therefore our greater values of $\Upsilon$ are not because of the resonances, but due to the background values of collision strengths ($\Omega_B$), which are $\sim$10$^{-3}$ and so are the $\Upsilon$ results. We also note that at energies above thresholds (9--30~Ryd range) our values of $\Omega$ are comparable with the DW results of \cite{ble}, and therefore we have confidence in our data. In conclusion, for some transitions for which our $\Upsilon$ are larger than of \cite{sst}, the differences are (probably) due to the corresponding values of  $\Omega_B$ -- see also section~1 regarding the discussion about the 5--10 transition.

\section{Conclusions}

In this paper we have reported energies and lifetimes for the lowest 86 levels of the 2s$^2$2p$^4$, 2s2p$^5$, 2p$^6$, and 2s$^2$2p$^3$3$\ell$ configurations of Mg~V. Differences with the compiled experimental energies of NIST are within 0.1~Ryd for most levels, except 2p$^6$~$^1$S$_0$ for which our result is higher by 0.4~Ryd. In addition, the ordering of levels also slightly differs in a few instances. Similarly, there is a good agreement (within 20\%) between our calculated values of $\tau$ and those of \cite{tff} (available on the web) for most levels, and differences for a few are within a factor of two, due to the corresponding differences in the A-values. Radiative rates for E1. E2, M1, and M2 transitions are also provided and there is a good agreement with other  available theoretical results for E1 transitions. Although scope remains for improvement in the accuracy of the reported data, due to computational restraints it is not possible in the present work, bearing in mind our further calculations for the more important collisional parameters.

Collision strengths  have been calculated with the fully relativistic {\sc darc} code and are listed for all  transitions,  from ground to higher excited levels,  at energies up to 30~Ryd. Furthermore, resonances in the thresholds region have been resolved in a fine energy mesh (0.0001~Ryd) to determine $\Upsilon$ values at temperatures up to 10$^6$~K, sufficient for applications to the modelling of a variety of plasmas. Similar results for the same number of transitions and with the $R$-matrix code (BSR) are available \cite{sst}, but discrepancies between the two sets of data are significant (up to two orders of magnitude) for over 60\% of the  transitions, and at all temperatures. In some cases the earlier $\Upsilon$ are larger,  but for a majority of transitions our results are higher. These discrepancies have been discussed in detail and the likely reasons for them include:  pseudo resonances, overestimation in the values of $\Omega$ at higher energies, and incorrect trends in the behaviour of transitions. Based on a number of comparisons, the earlier values of $\Upsilon$ by \cite{sst} are assessed to be inaccurate.

 Some discrepancies between the present and \cite{sst} sets of $\Upsilon$ 
are expected because the atomic structures are different. However, the scale of the discrepancies cannot be explained by differences in the atomic structure alone. For example, discrepancies in $\Upsilon$ for the allowed transitions are not proportionate to the corresponding differences in the f-values. Similarly, for many forbidden transitions (including the weak(er) ones), the $\Upsilon$  trends of \cite{sst} cannot be correct, because their results significantly increase with increasing temperature whereas the reverse is expected. 

Our presented results for $\Upsilon$ are probably the best available todate and therefore, we believe,  should be adopted for the modelling and diagnostics of plasmas. However, scope remains for improvement. Apart from improving the accuracy of the wavefunctions, the main scope is in expanding the collisional calculations to all 226 levels of Mg~V. This will definitely improve the accuracy of transitions, particularly among the higher levels, because all remaining 140 levels lie just above the lowest 86 and at energies below 13.7 ~Ryd. Resonances arising from these levels will significantly contribute to the calculations of $\Upsilon$. Unfortunately our present computational resources are inadequate to undertake such a large calculation, but that may be possible sometime in the future.



\newpage
\clearpage

\begin{flushleft}
Table 1. Comparison of energy levels (in Ryd) of Mg~V and their lifetimes (s).  $a{\pm}b \equiv a{\times}$10$^{{\pm}b}$.
\end{flushleft}
\begin{tabular}{rllrrrrrrl} \hline
Index  & Configuration       & Level    &  NIST     &   GRASP  &  MCHF1     & MCHF2   & SS       &$\tau$ (GRASP)& $\tau$ (MCHF2) \\
 \hline
    1  &  2s$^2$2p$^4$      &  $^3$P$  _2$    & 0.00000   & 0.00000  &  0.00000   & 0.00000 & 0.00000  &  .....       & .....       \\
    2  &  2s$^2$2p$^4$      &  $^3$P$  _1$    & 0.01625   & 0.01580  &  0.01577   & 0.01621 & 0.01729  &  8.541$+$00  & 7.916$+$00  \\
    3  &  2s$^2$2p$^4$      &  $^3$P$  _0$    & 0.02298   & 0.02237  &  0.02285   & 0.02285 & 0.02441  &  4.948$+$01  & 4.799$+$01  \\
    4  &  2s$^2$2p$^4$      &  $^1$D$  _2$    & 0.32738   & 0.35720  &  0.32932   & 0.32740 & 0.35311  &  4.006$-$01  & 4.110$-$01  \\
    5  &  2s$^2$2p$^4$      &  $^1$S$  _0$    & 0.70422   & 0.70631  &  0.70649   & 0.70422 & 0.68326  &  4.248$-$02  & 3.830$-$02  \\
    6  &  2s2p$^5$          &  $^3$P$^o_2$    & 2.58082   & 2.67741  &  2.59554   & 2.58092 & 2.70103  &  8.844$-$11  & 1.194$-$10  \\
    7  &  2s2p$^5$          &  $^3$P$^o_1$    & 2.59555   & 2.69165  &  2.61012   & 2.59560 & 2.71721  &  8.768$-$11  & 1.183$-$10  \\
    8  &  2s2p$^5$          &  $^3$P$^o_0$    & 2.60360   & 2.69938  &  2.61755   & 2.60361 & 2.72595  &  8.731$-$11  & 1.178$-$10  \\
    9  &  2s2p$^5$          &  $^1$P$^o_1$    & 3.62212   & 3.82766  &  3.63373   & 3.62219 & 3.87719  &  2.242$-$11  & 3.008$-$11  \\
   10  &  2p$^3$3s          &  $^5$S$^o_2$    &           & 5.96595  &  6.09758   & 6.09411 & 5.97559  &  1.641$-$07  & 2.521$-$07  \\
   11  &  2p$^3$3s          &  $^3$S$^o_1$    & 6.23800   & 6.11614  &  6.24023   & 6.23800 & 6.13389  &  1.834$-$11  & 2.092$-$11  \\
   12  &  2p$^6$            &  $^1$S$  _0$    & 6.04143   & 6.40649  &  6.04229   & 6.04143 & 6.55688  &  3.017$-$11  & 4.434$-$11  \\
   13  &  2p$^3$3s          &  $^3$D$^o_1$    & 6.63204   & 6.53478  &  6.63926   & 6.63202 & 6.53751  &  6.022$-$11  & 6.439$-$11  \\
   14  &  2p$^3$3s          &  $^3$D$^o_2$    & 6.63186   & 6.53457  &  6.63889   & 6.63182 & 6.53718  &  6.030$-$11  & 6.435$-$11  \\
   15  &  2p$^3$3s          &  $^3$D$^o_3$    & 6.63167   & 6.53453  &  6.63853   & 6.63167 & 6.53674  &  6.011$-$11  & 6.403$-$11  \\
   16  &  2p$^3$3p          &  $^5$P$  _1$    &           & 6.54144  &  6.67232   & 6.66804 & 6.54579  &  1.842$-$09  & 2.001$-$09  \\
   17  &  2p$^3$3p          &  $^5$P$  _2$    &           & 6.54254  &  6.67315   & 6.66915 & 6.54692  &  1.833$-$09  & 1.991$-$09  \\
   18  &  2p$^3$3p          &  $^5$P$  _3$    &           & 6.54439  &  6.67446   & 6.67100 & 6.55157  &  1.814$-$09  & 1.971$-$09  \\
   19  &  2p$^3$3s          &  $^1$D$^o_2$    & 6.70279   & 6.60924  &  6.70919   & 6.70280 & 6.61598  &  2.289$-$11  & 2.541$-$11  \\
   20  &  2p$^3$($^4$S)3p   &  $^3$P$  _1$    &           & 6.69317  &  6.79857   & 6.79791 & 6.70249  &  1.691$-$09  & 2.043$-$09  \\
   21  &  2p$^3$($^4$S)3p   &  $^3$P$  _0$    &           & 6.69366  &  6.79888   & 6.79845 & 6.70315  &  1.695$-$09  & 2.041$-$09  \\
   22  &  2p$^3$($^4$S)3p   &  $^3$P$  _2$    &           & 6.69375  &  6.79892   & 6.79822 & 6.70255  &  1.685$-$09  & 2.040$-$09  \\
   23  &  2p$^3$3s          &  $^3$P$^o_0$    & 6.89415   & 6.76192  &  6.91995   & 6.89415 & 6.88675  &  6.402$-$11  & 6.301$-$11  \\
   24  &  2p$^3$3s          &  $^3$P$^o_1$    & 6.89434   & 6.76225  &  6.92013   & 6.89440 & 6.88689  &  6.390$-$11  & 6.271$-$11  \\
   25  &  2p$^3$3s          &  $^3$P$^o_2$    & 6.89502   & 6.76325  &  6.92058   & 6.89518 & 6.88737  &  6.314$-$11  & 6.186$-$11  \\
   26  &  2p$^3$3s          &  $^1$P$^o_1$    & 6.96780   & 6.84056  &  6.97944   & 6.96782 & 6.96662  &  2.464$-$11  & 2.694$-$11  \\
   27  &  2p$^3$($^2$D)3p   &  $^1$P$  _1$    &           & 7.04952  &  7.16343   & 7.14737 & 7.06445  &  1.443$-$09  & 2.420$-$09  \\
   28  &  2p$^3$($^2$D)3p   &  $^3$D$  _2$    &           & 7.06809  &  7.17472   & 7.16871 & 7.07235  &  1.743$-$09  & 2.142$-$09  \\
   29  &  2p$^3$($^2$D)3p   &  $^3$D$  _1$    &           & 7.06847  &  7.17618   & 7.16876 & 7.07444  &  1.679$-$09  & 2.152$-$09  \\
   30  &  2p$^3$($^2$D)3p   &  $^3$D$  _3$    &           & 7.07096  &  7.17714   & 7.17130 & 7.07409  &  1.678$-$09  & 2.087$-$09  \\
   31  &  2p$^3$3p          &  $^3$F$  _2$    &           & 7.10886  &  7.23025   & 7.21029 & 7.09866  &  1.749$-$09  & 1.862$-$09  \\
   32  &  2p$^3$3p          &  $^3$F$  _3$    &           & 7.10996  &  7.23091   & 7.21135 & 7.09957  &  1.738$-$09  & 1.852$-$09  \\
   33  &  2p$^3$3p          &  $^3$F$  _4$    &           & 7.11132  &  7.23194   & 7.21268 & 7.10062  &  1.743$-$09  & 1.842$-$09  \\
   34  &  2p$^3$3p          &  $^1$F$  _3$    &           & 7.13452  &  7.25519   & 7.23482 & 7.12800  &  2.289$-$09  & 2.342$-$09  \\
   35  &  2p$^3$3d          &  $^5$D$^o_0$    &           & 7.26862  &  7.39656   & 7.39400 & 7.26697  &  7.173$-$10  & 7.831$-$10  \\
   36  &  2p$^3$3d          &  $^5$D$^o_1$    &           & 7.26857  &  7.39654   & 7.39396 & 7.26695  &  7.200$-$10  & 7.858$-$10  \\
   37  &  2p$^3$3d          &  $^5$D$^o_2$    &           & 7.26849  &  7.39651   & 7.39390 & 7.26690  &  7.258$-$10  & 7.915$-$10  \\
   38  &  2p$^3$3d          &  $^5$D$^o_3$    &           & 7.26839  &  7.39647   & 7.39382 & 7.26687  &  7.334$-$10  & 7.990$-$10  \\
   39  &  2p$^3$3d          &  $^5$D$^o_4$    &           & 7.26833  &  7.39641   & 7.39378 & 7.26691  &  7.378$-$10  & 8.033$-$10  \\
   40  &  2p$^3$($^2$P)3p   &  $^3$S$  _1$    &           & 7.28242  &  7.43539   & 7.42656 & 7.42244  &  2.731$-$10  & 5.231$-$10  \\
   41  &  2p$^3$($^2$D)3p   &  $^3$P$  _0$    &           & 7.29234  &  7.38555   & 7.35949 & 7.31579  &  2.702$-$10  & 4.333$-$10  \\
   42  &  2p$^3$($^2$D)3p   &  $^3$P$  _2$    &           & 7.29832  &  7.38764   & 7.35832 & 7.31740  &  2.501$-$10  & 4.315$-$10  \\
   43  &  2p$^3$($^2$D)3p   &  $^3$P$  _1$    &           & 7.30377  &  7.38479   & 7.35844 & 7.32139  &  2.684$-$10  & 4.228$-$10  \\
   44  &  2p$^3$($^2$P)3p   &  $^3$D$  _2$    &           & 7.33482  &  7.47484   & 7.47095 & 7.38883  &  6.843$-$10  & 1.014$-$09  \\
   45  &  2p$^3$($^2$P)3p   &  $^3$D$  _1$    &           & 7.33626  &  7.47311   & 7.46902 & 7.38865  &  6.850$-$10  & 1.045$-$09  \\
 \hline  											      
\end{tabular}   
\newpage
\clearpage

\begin{flushleft}
Table 1. Comparison of energy levels (in Ryd) of Mg~V and their lifetimes (s).  $a{\pm}b \equiv a{\times}$10$^{{\pm}b}$.
\end{flushleft}
\begin{tabular}{rllrrrrrrl} \hline
Index  & Configuration       & Level    &  NIST     &   GRASP  &  MCHF1     & MCHF2   & SS       &$\tau$ (GRASP)& $\tau$ (MCHF2) \\
 \hline
   46  &  2p$^3$($^2$P)3p   &  $^3$D$  _3$   &           & 7.33629  &  7.47286   & 7.46937 & 7.38865  &  6.968$-$10  & 1.055$-$09  \\
   47  &  2p$^3$($^2$P)3p   &  $^1$D$  _2$   &           & 7.35689  &  7.45450   & 7.44935 & 7.40625  &  6.284$-$10  & 7.320$-$10  \\
   48  &  2p$^3$($^2$P)3p   &  $^1$P$  _1$   &           & 7.37851  &  7.51281   & 7.50487 & 7.48245  &  8.614$-$10  & 1.207$-$09  \\
   49  &  2p$^3$($^4$S)3d   &  $^3$D$^o_1$   &  7.49038  & 7.38496  &  7.49488   & 7.49039 & 7.44750  &  1.042$-$11  & 1.404$-$11  \\
   50  &  2p$^3$($^4$S)3d   &  $^3$D$^o_2$   &  7.49052  & 7.38517  &  7.49504   & 7.49055 & 7.44880  &  1.036$-$11  & 1.395$-$11  \\
   51  &  2p$^3$($^4$S)3d   &  $^3$D$^o_3$   &  7.49122  & 7.38607  &  7.49571   & 7.49131 & 7.44868  &  1.018$-$11  & 1.371$-$11  \\
   52  &  2p$^3$($^2$P)3p   &  $^3$P$  _2$   &           & 7.42960  &  7.56479   & 7.54766 & 7.56987  &  7.407$-$10  & 1.104$-$09  \\
   53  &  2p$^3$($^2$P)3p   &  $^3$P$  _1$   &           & 7.43539  &  7.56778   & 7.54814 & 7.57590  &  6.347$-$10  & 1.056$-$09  \\
   54  &  2p$^3$($^2$P)3p   &  $^3$P$  _0$   &           & 7.43817  &  7.56941   & 7.54830 & 7.57869  &  5.933$-$10  & 1.028$-$09  \\
   55  &  2p$^3$($^2$D)3p   &  $^1$D$  _2$   &           & 7.57783  &  7.64766   & 7.62243 & 7.67477  &  4.665$-$10  & 7.756$-$10  \\
   56  &  2p$^3$($^2$D)3d   &  $^3$F$^o_2$   &           & 7.79032  &  7.89914   & 7.89138 & 7.78774  &  6.667$-$10  & 7.450$-$10  \\
   57  &  2p$^3$($^2$D)3d   &  $^3$F$^o_3$   &           & 7.79268  &  7.90120   & 7.89340 & 7.78911  &  6.450$-$10  & 7.185$-$10  \\
   58  &  2p$^3$($^2$D)3d   &  $^1$S$^o_0$   &           & 7.79486  &  7.90496   & 7.89451 & 7.79068  &  6.405$-$10  & 7.166$-$10  \\
   59  &  2p$^3$($^2$D)3d   &  $^3$F$^o_4$   &           & 7.79565  &  7.90376   & 7.89600 & 7.79092  &  7.759$-$10  & 8.492$-$10  \\
   60  &  2p$^3$($^2$P)3p   &  $^1$S$  _0$   &           & 7.81709  &  7.84102   & 7.83254 & 7.97019  &  1.989$-$10  & 3.358$-$10  \\
   61  &  2p$^3$($^2$D)3d   &  $^3$G$^o_3$   &           & 7.82605  &  7.93922   & 7.92330 & 7.81099  &  6.703$-$10  & 7.482$-$10  \\
   62  &  2p$^3$($^2$D)3d   &  $^3$G$^o_4$   &           & 7.82591  &  7.93903   & 7.92317 & 7.81076  &  7.396$-$10  & 8.111$-$10  \\
   63  &  2p$^3$($^2$D)3d   &  $^3$G$^o_5$   &           & 7.82562  &  7.93884   & 7.92286 & 7.81039  &  7.441$-$10  & 8.162$-$10  \\
   64  &  2p$^3$($^2$D)3d   &  $^1$G$^o_4$   &           & 7.83854  &  7.94236   & 7.93125 & 7.82390  &  7.772$-$10  & 8.557$-$10  \\
   65  &  2p$^3$($^2$D)3d   &  $^3$D$^o_1$   &  7.94069  & 7.85135  &  7.94646   & 7.94073 & 7.84904  &  1.122$-$11  & 1.450$-$11  \\
   66  &  2p$^3$($^2$D)3d   &  $^3$D$^o_3$   &  7.93910  & 7.85077  &  7.94502   & 7.93908 & 7.84889  &  9.563$-$12  & 1.271$-$11  \\
   67  &  2p$^3$($^2$D)3d   &  $^3$D$^o_2$   &  7.94039  & 7.85145  &  7.94613   & 7.94031 & 7.84926  &  1.043$-$11  & 1.366$-$11  \\
   68  &  2p$^3$($^2$D)3d   &  $^1$P$^o_1$   &  7.95952  & 7.87079  &  7.96346   & 7.96449 & 7.87547  &  9.325$-$12  & 1.078$-$11  \\
   69  &  2p$^3$($^2$D)3d   &  $^1$D$^o_2$   &  7.99738  & 7.90109  &  8.01933   & 7.99921 & 7.91894  &  7.758$-$12  & 1.108$-$11  \\
   70  &  2p$^3$($^2$D)3d   &  $^3$P$^o_2$   &  7.98994  & 7.90223  &  8.00795   & 7.99003 & 7.90587  &  6.390$-$12  & 6.608$-$12  \\
   71  &  2p$^3$($^2$D)3d   &  $^3$P$^o_1$   &  7.99439  & 7.90700  &  8.01206   & 7.99446 & 7.90851  &  5.784$-$12  & 6.994$-$12  \\
   72  &  2p$^3$($^2$D)3d   &  $^3$P$^o_0$   &  7.99603  & 7.90912  &  8.01369   & 7.99605 & 7.90955  &  6.068$-$12  & 7.202$-$12  \\
   73  &  2p$^3$($^2$D)3d   &  $^3$S$^o_1$   &  8.01473  & 7.91670  &  8.02968   & 8.01475 & 7.91837  &  5.202$-$12  & 6.563$-$12  \\
   74  &  2p$^3$($^2$D)3d   &  $^1$F$^o_3$   &  8.04458  & 7.97597  &  8.05239   & 8.04461 & 7.97590  &  6.031$-$12  & 7.294$-$12  \\
   75  &  2p$^3$($^2$P)3d   &  $^3$F$^o_4$   &           & 8.05087  &  8.19519   & 8.18314 & 8.16058  &  7.172$-$10  & 7.796$-$10  \\
   76  &  2p$^3$($^2$P)3d   &  $^3$F$^o_3$   &           & 8.05381  &  8.19779   & 8.18557 & 8.16238  &  3.274$-$10  & 4.828$-$10  \\
   77  &  2p$^3$($^2$P)3d   &  $^3$F$^o_2$   &           & 8.05551  &  8.19942   & 8.18716 & 8.16354  &  4.830$-$10  & 4.865$-$10  \\
   78  &  2p$^3$($^2$P)3d   &  $^3$P$^o_0$   &  8.19008  & 8.05591  &  8.21337   & 8.19009 & 8.17823  &  2.627$-$11  & 2.573$-$11  \\
   79  &  2p$^3$($^2$P)3d   &  $^3$P$^o_1$   &  8.19195  & 8.05827  &  8.21523   & 8.19198 & 8.17931  &  3.061$-$11  & 2.890$-$11  \\
   80  &  2p$^3$($^2$P)3d   &  $^3$P$^o_2$   &  8.19565  & 8.06284  &  8.21863   & 8.19567 & 8.18148  &  5.029$-$11  & 4.041$-$11  \\
   81  &  2p$^3$($^2$P)3d   &  $^3$D$^o_2$   &  8.22427  & 8.09547  &  8.24650   & 8.21568 & 8.21649  &  1.084$-$11  & 1.467$-$11  \\
   82  &  2p$^3$($^2$P)3d   &  $^3$D$^o_1$   &  8.22661  & 8.09737  &  8.24425   & 8.21978 & 8.21441  &  1.012$-$11  & 1.373$-$11  \\
   83  &  2p$^3$($^2$P)3d   &  $^3$D$^o_3$   &  8.22101  & 8.09740  &  8.24482   & 8.22089 & 8.21344  &  1.256$-$11  & 1.654$-$11  \\
   84  &  2p$^3$($^2$P)3d   &  $^1$D$^o_2$   &  8.21484  & 8.11375  &  8.23410   &         & 8.20806  &  8.434$-$12  & .....       \\
   85  &  2p$^3$($^2$P)3d   &  $^1$F$^o_3$   &  8.25034  & 8.13524  &  8.27297   & 8.25038 & 8.25501  &  6.906$-$12  & 1.021$-$11  \\
   86  &  2p$^3$($^2$P)3d   &  $^1$P$^o_1$   &  8.33354  & 8.23226  &  8.35818   & 8.33354 & 8.35849  &  4.140$-$12  & 5.735$-$12  \\
 \hline  											      
\end{tabular}   								   					       
			      							   					       
\vspace*{0.5 cm}													       
\begin{flushleft}													       
{\small
NIST: {\tt http://www.nist.gov/pml/data/asd.cfm} \\
GRASP: Energies from the {\sc grasp} code for 226 level calculations \\
MCHF1: Energies of Tayal and Sossah \cite{sst} from the MCHF code  \\ 
MCHF2: Energies and liftimes of Tachiev and Froese Fischer \cite{tff} from the MCHF code, downloaded on 21 March 2016 from {\tt  http://nlte.nist.gov/MCHF/view.html}  \\
SS: Energies of Bhatia et al. \cite{ble} from the SS code \\	
}															       
\end{flushleft} 

\vspace{2.0 cm} 
\begin{flushleft}
Table 2. Transition wavelengths ($\lambda_{ij}$ in $\rm \AA$), radiative rates (A$_{ji}$ in s$^{-1}$), oscillator strengths (f$_{ij}$, dimensionless), and line  
strengths (S, in atomic units) for electric dipole (E1), and A$_{ji}$ for E2, M1, and M2 transitions in Mg V.  Ratio (R) of the velocity and length forms of 
A-values for E1 transitions is also listed in the last column.  $a{\pm}b \equiv a{\times}$10$^{{\pm}b}$. See Table 1 for level indices. Complete table is available online as Supporting Information.
\end{flushleft}
\begin{tabular}{rrrrrrrrrr}                                                                                                                                      
\hline                                                                                                                                                                                                                                                                                                               
$i$ & $j$ & $\lambda_{ij}$ & A$^{{\rm E1}}_{ji}$  & f$^{{\rm E1}}_{ij}$ & S$^{{\rm E1}}$ & A$^{{\rm E2}}_{ji}$  & A$^{{\rm M1}}_{ji}$ & A$^{{\rm M2}}_{ji}$  & R  \\  
\hline                                                                                                                                                   
     1 &    2 &  5.767$+$04 &  0.000$+$00 &  0.000$+$00 &  0.000$+$00 &  1.184$-$07 &  1.171$-$01 &  0.000$+$00 &  0.000$+$00 \\             
    1 &    3 &  4.073$+$04 &  0.000$+$00 &  0.000$+$00 &  0.000$+$00 &  9.017$-$07 &  0.000$+$00 &  0.000$+$00 &  0.000$+$00 \\             
    1 &    4 &  2.551$+$03 &  0.000$+$00 &  0.000$+$00 &  0.000$+$00 &  2.157$-$03 &  1.923$+$00 &  0.000$+$00 &  0.000$+$00 \\             
    1 &    5 &  1.290$+$03 &  0.000$+$00 &  0.000$+$00 &  0.000$+$00 &  1.816$-$02 &  0.000$+$00 &  0.000$+$00 &  0.000$+$00 \\             
    1 &    6 &  3.404$+$02 &  8.517$+$09 &  1.479$-$01 &  8.287$-$01 &  0.000$+$00 &  0.000$+$00 &  6.560$+$00 &  7.900$-$01 \\             
    1 &    7 &  3.385$+$02 &  4.822$+$09 &  4.972$-$02 &  2.770$-$01 &  0.000$+$00 &  0.000$+$00 &  5.756$-$03 &  7.900$-$01 \\             
    1 &    8 &  3.376$+$02 &  0.000$+$00 &  0.000$+$00 &  0.000$+$00 &  0.000$+$00 &  0.000$+$00 &  4.226$+$00 &  0.000$+$00 \\             
    1 &    9 &  2.381$+$02 &  3.366$+$07 &  1.716$-$04 &  6.726$-$04 &  0.000$+$00 &  0.000$+$00 &  3.756$+$01 &  5.600$-$01 \\             
    1 &   10 &  1.527$+$02 &  4.779$+$06 &  1.672$-$05 &  4.203$-$05 &  0.000$+$00 &  0.000$+$00 &  3.219$+$01 &  8.900$-$01 \\      
    ... & \\
    ... & \\
    ... & \\
\hline                                                                                                                                                          
\end{tabular}     

\newpage
\clearpage
\begin{flushleft}
Table 3. Comparison of f-values for E1 transitions among the lowest 13 levels of Mg~V.  $a{\pm}b \equiv a{\times}$10$^{{\pm}b}$.   See Table 1 for level indices.
\end{flushleft}
\begin{tabular}{rrrrrrl} \hline
I & J& f (GRASP) & f (MCHF1) &  R  \\
\hline
     1  &  6    &  1.4792$-$01 &  1.192$-$1 & 7.9$-$01  \\
     1  &  7    &  4.9715$-$02 &  4.005$-$2 & 7.9$-$01  \\
     1  &  9    &  1.7163$-$04 &  1.677$-$4 & 5.6$-$01  \\
     1  & 10    &  1.6717$-$05 &  1.462$-$5 & 8.9$-$01  \\
     1  & 11    &  6.1254$-$02 &  5.416$-$2 & 7.3$-$01  \\
     1  & 13    &  1.0671$-$03 &  9.241$-$4 & 7.4$-$01  \\
     1  & 14    &  1.4640$-$02 &  1.278$-$2 & 7.3$-$01  \\
     1  & 15    &  6.7861$-$02 &  6.104$-$2 & 7.1$-$01  \\
     2  &  6    &  8.1588$-$02 &  6.572$-$2 & 7.9$-$01  \\
     2  &  7    &  4.9295$-$02 &  3.972$-$2 & 7.9$-$01  \\
     2  &  8    &  6.5997$-$02 &  5.317$-$2 & 7.9$-$01  \\
     2  &  9    &  6.2735$-$06 &  6.039$-$6 & 9.5$-$01  \\
     2  & 10    &  7.6740$-$06 &  7.179$-$6 & 9.2$-$01  \\
     2  & 11    &  5.9890$-$02 &  5.304$-$2 & 7.3$-$01  \\
     2  & 13    &  2.1725$-$02 &  1.934$-$2 & 7.2$-$01  \\
     2  & 14    &  5.6433$-$02 &  5.150$-$2 & 7.1$-$01  \\
     3  &  7    &  1.9643$-$01 &  1.502$-$1 & 7.9$-$01  \\
     3  &  9    &  3.9402$-$05 &  3.369$-$5 & 2.3$-$00  \\
     3  & 11    &  5.9666$-$02 &  5.278$-$2 & 7.3$-$01  \\
     3  & 13    &  7.5063$-$02 &  6.861$-$2 & 7.1$-$01  \\
     4  &  6    &  9.6461$-$05 &  9.680$-$5 & 7.0$-$01  \\
     4  &  7    &  5.3585$-$06 &  4.041$-$6 & 1.8$-$00  \\
     4  &  9    &  2.6098$-$01 &  2.178$-$1 & 5.5$-$01  \\
     4  & 10    &  1.3670$-$08 &       $ $  & 4.9$-$01  \\
     4  & 11    &  1.3852$-$05 &  1.728$-$5 & 6.0$-$01  \\
     4  & 13    &  7.2791$-$05 &  4.927$-$5 & 9.6$-$01  \\
     4  & 14    &  1.0104$-$05 &  1.122$-$5 & 7.5$-$01  \\
     4  & 15    &  5.6661$-$05 &  6.194$-$5 & 7.6$-$01  \\
     5  &  7    &  8.0804$-$05 &  6.419$-$5 & 4.1$-$01  \\
     5  &  9    &  9.5261$-$02 &  8.068$-$2 & 1.8$-$00  \\
     5  & 11    &  1.8733$-$06 &  6.363$-$7 & 3.9$-$01  \\
     5  & 13    &  2.0181$-$04 &  1.094$-$4 & 6.6$-$01  \\
     7  & 12    &  2.5105$-$05 &  2.398$-$5 & 7.5$-$01  \\
     9  & 12    &  2.0679$-$01 &  1.622$-$1 & 6.0$-$01  \\
    11  & 12    &  5.8370$-$09 &       $ $  & 5.3$-$01  \\
    12  & 13    &  3.2235$-$11 &  2.102$-$8 & 3.1$-$05  \\
\hline  
\end{tabular}

\begin{flushleft}
{\small
GRASP: Present calculations with the {\sc grasp} code \\
MCHF1: Calculations of \cite{sst} with the {\sc mchf} code \\
R: ratio of  the velocity and length forms of GRASP f-values \\
}
\end{flushleft}


\newpage
\clearpage
\begin{flushleft}
Table 4. Collision strengths for  all transitions from ground to higher excited levels of  Mg V. $a{\pm}b \equiv$ $a\times$10$^{{\pm}b}$. See Table 1 for level indices. Complete table is available online as Supporting Information.
\end{flushleft}
\begin{tabular}{rrlllllllr}                                                                                   
\hline                                                                                                                                                                                                             
\multicolumn{2}{c}{Transition} & \multicolumn{8}{c}{Energy (Ryd)}\\                                           
\hline                                                                                                        
  $i$ & $j$ &  9.0   &    12.0   &    15.0   &    18.0   &    21.0    &   24.0  &     27.0  &     30.0    \\                                                                
\hline 
    1   &   2  &  2.979$-$01  &  2.666$-$01  &  2.416$-$01  &  2.215$-$01  &  2.061$-$01  &  1.942$-$01  &  1.850$-$01  &  1.779$-$01 \\
    1   &   3  &  7.193$-$02  &  6.743$-$02  &  6.433$-$02  &  6.217$-$02  &  6.079$-$02  &  5.993$-$02  &  5.944$-$02  &  5.921$-$02 \\
    1   &   4  &  2.957$-$01  &  2.435$-$01  &  1.994$-$01  &  1.635$-$01  &  1.354$-$01  &  1.133$-$01  &  9.583$-$02  &  8.193$-$02 \\
    1   &   5  &  3.824$-$02  &  3.029$-$02  &  2.379$-$02  &  1.874$-$02  &  1.491$-$02  &  1.201$-$02  &  9.780$-$03  &  8.064$-$03 \\
    1   &   6  &  2.977$+$00  &  3.271$+$00  &  3.498$+$00  &  3.710$+$00  &  3.889$+$00  &  4.051$+$00  &  4.194$+$00  &  4.322$+$00 \\
    1   &   7  &  9.949$-$01  &  1.092$+$00  &  1.168$+$00  &  1.238$+$00  &  1.298$+$00  &  1.352$+$00  &  1.399$+$00  &  1.442$+$00 \\
    1   &   8  &  9.052$-$03  &  7.375$-$03  &  6.072$-$03  &  5.084$-$03  &  4.312$-$03  &  3.703$-$03  &  3.214$-$03  &  2.818$-$03 \\
    1   &   9  &  5.518$-$02  &  4.967$-$02  &  4.202$-$02  &  3.601$-$02  &  3.120$-$02  &  2.734$-$02  &  2.421$-$02  &  2.164$-$02 \\
    1   &  10  &  3.036$-$02  &  2.103$-$02  &  1.514$-$02  &  1.142$-$02  &  8.923$-$03  &  7.169$-$03  &  5.890$-$03  &  4.926$-$03 \\
    ... & \\
    ... & \\
    ... & \\
\hline                                                                                                        
\end{tabular}               

\newpage
\clearpage
\begin{flushleft}
Table 5. Effective collision strengths for transitions in  Mg V. $a{\pm}b \equiv a{\times}10^{{\pm}b}$. See Table 1 for level indices. Complete table is available online as Supporting Information.
\end{flushleft}
{\small
\begin{tabular}{rrlllllllllll}                                                                                                              
\hline                                                                                                                                      
\multicolumn {2}{c}{Transition} & \multicolumn{11}{c}{Temperature (log T$_e$, K)}\\                                                         
\hline                                                                                                                                      
$i$ & $j$ &     4.0  &    4.2  &    4.4   &   4.6  &    4.8  &    5.0  &   5.2  &    5.4 &  5.6  &  5.8  &  6.0  \\                         
\hline                                                                                                                                                        
    1 &    2 &  8.541$-$1 &  8.841$-$1 &  8.973$-$1 &  9.364$-$1 &  9.914$-$1 &  1.021$-$0 &  1.005$-$0 &  9.457$-$1 &  8.548$-$1 &  7.451$-$1 &  6.281$-$1 \\
    1 &    3 &  2.180$-$1 &  2.264$-$1 &  2.320$-$1 &  2.486$-$1 &  2.698$-$1 &  2.795$-$1 &  2.714$-$1 &  2.495$-$1 &  2.202$-$1 &  1.885$-$1 &  1.576$-$1 \\
    1 &    4 &  6.724$-$1 &  6.795$-$1 &  6.862$-$1 &  6.808$-$1 &  6.657$-$1 &  6.477$-$1 &  6.288$-$1 &  6.074$-$1 &  5.770$-$1 &  5.299$-$1 &  4.647$-$1 \\
    1 &    5 &  1.185$-$1 &  1.071$-$1 &  9.745$-$2 &  9.112$-$2 &  8.799$-$2 &  8.647$-$2 &  8.494$-$2 &  8.248$-$2 &  7.825$-$2 &  7.137$-$2 &  6.190$-$2 \\
    1 &    6 &  2.099$-$0 &  2.106$-$0 &  2.118$-$0 &  2.139$-$0 &  2.170$-$0 &  2.218$-$0 &  2.292$-$0 &  2.400$-$0 &  2.539$-$0 &  2.707$-$0 &  2.892$-$0 \\
    1 &    7 &  7.004$-$1 &  7.026$-$1 &  7.071$-$1 &  7.147$-$1 &  7.257$-$1 &  7.422$-$1 &  7.673$-$1 &  8.030$-$1 &  8.487$-$1 &  9.046$-$1 &  9.659$-$1 \\
    1 &    8 &  1.301$-$2 &  1.308$-$2 &  1.315$-$2 &  1.321$-$2 &  1.326$-$2 &  1.340$-$2 &  1.377$-$2 &  1.422$-$2 &  1.421$-$2 &  1.343$-$2 &  1.195$-$2 \\
    1 &    9 &  8.124$-$2 &  8.143$-$2 &  8.075$-$2 &  7.990$-$2 &  7.955$-$2 &  7.998$-$2 &  8.064$-$2 &  8.092$-$2 &  7.981$-$2 &  7.593$-$2 &  6.880$-$2 \\
    1 &   10 &  7.465$-$1 &  7.199$-$1 &  6.801$-$1 &  6.201$-$1 &  5.339$-$1 &  4.310$-$1 &  3.293$-$1 &  2.419$-$1 &  1.732$-$1 &  1.222$-$1 &  8.547$-$2 \\
    ... & \\
    ... & \\
    ... & \\
\hline                                                                                                                                      
\end{tabular}          
}													       

\newpage
\clearpage
\begin{flushleft}
Table 6. Oscillator strengths (f, dimensionless), collision strengths ($\Omega$), and effective collision strengths ($\Upsilon$) for transitions of  Mg V with upper level J=86. $a{\pm}b \equiv$ $a\times$10$^{{\pm}b}$.
\end{flushleft}
{\tiny
\begin{tabular}{rrllllllllllll}                                                                                   
\hline   
& & Present f &  {\bf f of TS15 } & \multicolumn{3}{c}{Present $\Omega$} &  \multicolumn{3}{c}{Present $\Upsilon$}  &  \multicolumn{3}{c}{$\Upsilon$ of TS15}     \\                                                                                                                                                                                                   
\multicolumn{2}{c}{Transition} & &  & \multicolumn{3}{c}{Energy (Ryd)} & \multicolumn{3}{c}{Temperature (log, K)} & \multicolumn{3}{c}{Temperature (log, K)}\\                                           
\hline                                                                                                        
  $i$ & $j$ &       &      & 9.0         &    12.0     &    15.0     &   4.0       &  5.0        &  6.0         &   4.0        &  5.0        &  6.0       \\                                                                
\hline               
    1   &  86 &   5.59$-$07 & 3.80$-$6 &  9.24$-$03  &  5.65$-$03  &  3.65$-$03  &  1.04$-$2  &  9.85$-$3  &  5.44$-$3   & 2.64$-$3  & 4.80$-$3 & 3.17$-$3 \\
    2   &  86 &   1.25$-$05 & 8.05$-$6 &  6.21$-$03  &  3.82$-$03  &  2.48$-$03  &  6.97$-$3  &  6.59$-$3  &  3.66$-$3   & 1.79$-$3  & 3.25$-$3 & 2.13$-$3 \\
    3   &  86 &   3.17$-$04 & 1.98$-$4 &  2.20$-$03  &  1.41$-$03  &  9.72$-$04  &  2.51$-$3  &  2.38$-$3  &  1.38$-$3   & 6.39$-$4  & 1.17$-$3 & 8.03$-$4 \\
    4   &  86 &   1.64$-$02 & 9.11$-$3 &  2.11$-$02  &  2.87$-$02  &  3.46$-$02  &  2.00$-$2  &  2.17$-$2  &  3.17$-$2   & 5.86$-$3  & 1.22$-$2 & 2.10$-$2 \\
    5   &  86 &   1.49$-$00 & 1.05$-$0 &  2.90$-$01  &  3.84$-$01  &  4.98$-$01  &  2.33$-$1  &  2.60$-$1  &  4.44$-$1   & 7.93$-$2  & 1.72$-$1 & 3.43$-$1 \\
    6   &  86 &   ........  &          &  1.35$-$02  &  7.86$-$04  &  4.78$-$04  &  2.86$-$2  &  2.47$-$2  &  7.66$-$3   & 5.08$-$4  & 9.30$-$4 & 5.92$-$4 \\
    7   &  86 &   ........  &          &  7.95$-$03  &  5.00$-$04  &  3.15$-$04  &  1.69$-$2  &  1.46$-$2  &  4.56$-$3   & 3.18$-$4  & 5.84$-$4 & 3.74$-$4 \\
    8   &  86 &   ........  &          &  2.60$-$03  &  1.63$-$04  &  1.01$-$04  &  5.51$-$3  &  4.77$-$3  &  1.48$-$3   & 1.07$-$4  & 1.97$-$4 & 1.26$-$4 \\
    9   &  86 &   ........  &          &  4.38$-$02  &  3.84$-$03  &  4.59$-$03  &  1.27$-$2  &  1.15$-$2  &  6.69$-$3   & 6.28$-$4  & 1.30$-$3 & 2.18$-$3 \\
   10   &  86 &   ........  &          &  3.37$-$06  &  1.11$-$06  &  6.70$-$07  &  5.31$-$6  &  4.75$-$6  &  1.89$-$6   & 9.90$-$7  & 1.58$-$6 & 1.35$-$6 \\
   11   &  86 &   ........  &          &  3.32$-$04  &  2.33$-$04  &  1.66$-$04  &  4.27$-$4  &  4.03$-$4  &  2.30$-$4   & 1.64$-$4  & 3.02$-$4 & 2.16$-$4 \\
   12   &  86 &   3.56$-$04 & 4.28$-$5 &  2.16$-$03  &  1.53$-$03  &  1.62$-$03  &  4.15$-$3  &  3.81$-$3  &  2.31$-$3   & 2.34$-$5  & 4.43$-$5 & 5.03$-$5 \\
   13   &  86 &   ........  &          &  2.26$-$03  &  6.10$-$04  &  3.15$-$04  &  7.78$-$3  &  6.79$-$3  &  2.25$-$3   & 6.13$-$4  & 1.01$-$3 & 4.99$-$4 \\
   14   &  86 &   ........  &          &  3.79$-$03  &  1.02$-$03  &  5.25$-$04  &  1.27$-$2  &  1.11$-$2  &  3.68$-$3   & 1.03$-$3  & 1.69$-$3 & 8.38$-$4 \\
   15   &  86 &   ........  &          &  5.40$-$03  &  1.43$-$03  &  7.38$-$04  &  1.73$-$2  &  1.51$-$2  &  5.04$-$3   & 1.45$-$3  & 2.38$-$3 & 1.17$-$3 \\
   16   &  86 &   3.10$-$13 &          &  6.69$-$06  &  1.97$-$06  &  1.09$-$06  &  1.35$-$5  &  1.19$-$5  &  4.33$-$6   & 2.04$-$6  & 3.30$-$6 & 2.36$-$6 \\
   17   &  86 &   1.12$-$10 &          &  9.90$-$06  &  1.90$-$06  &  1.12$-$06  &  1.11$-$5  &  9.91$-$6  &  3.74$-$6   & 2.14$-$6  & 3.63$-$6 & 4.95$-$6 \\
   18   &  86 &   ........  &          &  1.42$-$05  &  3.40$-$06  &  2.00$-$06  &  1.73$-$5  &  1.55$-$5  &  6.01$-$6   & 3.46$-$6  & 5.04$-$6 & 3.37$-$6 \\
   19   &  86 &   ........  &          &  2.60$-$02  &  2.86$-$02  &  3.08$-$02  &  8.96$-$2  &  8.16$-$2  &  4.53$-$2   & 7.57$-$3  & 1.57$-$2 & 2.10$-$2 \\
   20   &  86 &   1.09$-$09 & 2.66$-$7 &  2.10$-$03  &  7.78$-$04  &  4.45$-$04  &  1.66$-$3  &  1.55$-$3  &  7.79$-$4   & 6.55$-$4  & 1.16$-$3 & 6.77$-$4 \\
   21   &  86 &   1.49$-$05 & 2.83$-$6 &  7.69$-$04  &  3.41$-$04  &  2.45$-$04  &  6.15$-$4  &  5.83$-$4  &  3.48$-$4   & 2.22$-$4  & 3.98$-$4 & 2.56$-$4 \\
   22   &  86 &   2.17$-$08 & 2.08$-$7 &  3.35$-$03  &  1.28$-$03  &  7.27$-$04  &  2.57$-$3  &  2.41$-$3  &  1.24$-$3   & 1.11$-$3  & 1.96$-$3 & 1.12$-$3 \\
   23   &  86 &   ........  &          &  1.14$-$02  &  5.28$-$03  &  2.69$-$03  &  2.46$-$2  &  2.20$-$2  &  8.59$-$3   & 7.03$-$3  & 1.16$-$2 & 5.42$-$3 \\
   24   &  86 &   ........  &          &  3.49$-$02  &  1.61$-$02  &  8.24$-$03  &  7.33$-$2  &  6.56$-$2  &  2.58$-$2   & 2.13$-$2  & 3.50$-$2 & 1.64$-$2 \\
   25   &  86 &   ........  &          &  6.17$-$02  &  2.89$-$02  &  1.54$-$02  &  1.23$-$1  &  1.10$-$1  &  4.43$-$2   & 3.63$-$2  & 5.98$-$2 & 2.85$-$2 \\
   26   &  86 &   ........  &          &  5.47$-$01  &  7.33$-$01  &  8.07$-$01  &  4.28$-$1  &  4.81$-$1  &  6.96$-$1   & 2.62$-$1  & 5.60$-$1 & 8.06$-$1 \\
   27   &  86 &   7.34$-$04 & 1.40$-$3 &  2.09$-$02  &  1.69$-$02  &  1.85$-$02  &  3.79$-$2  &  3.53$-$2  &  2.35$-$2   & 7.77$-$3  & 1.52$-$2 & 2.02$-$2 \\
   28   &  86 &   3.43$-$06 & 2.28$-$6 &  1.98$-$02  &  2.14$-$03  &  1.16$-$03  &  6.60$-$2  &  5.72$-$2  &  1.78$-$2   & 2.31$-$3  & 3.80$-$3 & 1.89$-$3 \\
   29   &  86 &   7.95$-$05 & 4.04$-$5 &  1.33$-$02  &  3.71$-$03  &  3.38$-$03  &  3.71$-$2  &  3.25$-$2  &  1.21$-$2   & 3.22$-$3  & 5.90$-$3 & 6.18$-$3 \\
   30   &  86 &   ........  &          &  2.80$-$02  &  2.72$-$03  &  1.34$-$03  &  9.24$-$2  &  8.00$-$2  &  2.47$-$2   & 3.16$-$3  & 5.17$-$3 & 2.42$-$3 \\
   31   &  86 &   2.20$-$06 & 2.21$-$7 &  2.39$-$02  &  1.14$-$03  &  6.16$-$04  &  1.30$-$2  &  1.14$-$2  &  3.86$-$3   & 1.29$-$3  & 2.06$-$3 & 9.54$-$4 \\
   32   &  86 &   ........  &          &  3.50$-$02  &  1.49$-$03  &  7.29$-$04  &  1.48$-$2  &  1.30$-$2  &  4.44$-$3   & 1.80$-$3  & 2.89$-$3 & 1.30$-$3 \\
   33   &  86 &   ........  &          &  4.73$-$02  &  1.88$-$03  &  8.96$-$04  &  1.67$-$2  &  1.47$-$2  &  5.08$-$3   & 2.37$-$3  & 3.75$-$3 & 1.62$-$3 \\
   34   &  86 &   ........  &          &  3.68$-$02  &  1.64$-$02  &  1.56$-$02  &  4.28$-$2  &  3.94$-$2  &  2.25$-$2   & 6.76$-$3  & 1.29$-$2 & 1.28$-$2 \\
   35   &  86 &   ........  &          &  1.30$-$05  &  8.77$-$06  &  6.80$-$06  &  1.98$-$5  &  1.84$-$5  &  9.93$-$6   & 4.07$-$6  & 8.45$-$6 & 8.03$-$6 \\
   36   &  86 &   ........  &          &  3.50$-$05  &  2.32$-$05  &  1.79$-$05  &  5.12$-$5  &  4.77$-$5  &  2.60$-$5   & 1.22$-$5  & 2.40$-$5 & 2.36$-$5 \\
   37   &  86 &   ........  &          &  4.83$-$05  &  3.13$-$05  &  2.42$-$05  &  6.52$-$5  &  6.10$-$5  &  3.41$-$5   & 1.99$-$5  & 3.87$-$5 & 4.16$-$5 \\
   38   &  86 &   ........  &          &  5.93$-$05  &  3.94$-$05  &  3.10$-$05  &  7.61$-$5  &  7.17$-$5  &  4.18$-$5   & 2.08$-$5  & 3.79$-$5 & 5.85$-$5 \\
   39   &  86 &   ........  &          &  1.02$-$04  &  7.74$-$05  &  6.14$-$05  &  1.45$-$4  &  1.37$-$4  &  8.13$-$5   & 3.48$-$5  & 6.04$-$5 & 4.88$-$5 \\
   40   &  86 &   5.73$-$05 & 8.34$-$6 &  4.97$-$02  &  2.54$-$02  &  1.42$-$02  &  9.82$-$2  &  8.86$-$2  &  3.71$-$2   & 5.77$-$2  & 9.30$-$2 & 4.15$-$2 \\
   41   &  86 &   1.54$-$04 & 7.96$-$5 &  4.54$-$03  &  2.97$-$03  &  2.63$-$03  &  6.62$-$3  &  6.16$-$3  &  3.72$-$3   & 1.27$-$3  & 2.13$-$3 & 1.22$-$3 \\
   42   &  86 &   1.33$-$06 & 9.57$-$8 &  1.60$-$02  &  6.80$-$03  &  3.87$-$03  &  2.17$-$2  &  1.97$-$2  &  8.78$-$3   & 5.43$-$3  & 8.95$-$3 & 4.30$-$3 \\
   43   &  86 &   3.78$-$05 & 5.93$-$6 &  4.34$-$02  &  2.14$-$02  &  1.19$-$02  &  8.00$-$2  &  7.22$-$2  &  3.05$-$2   & 5.28$-$3  & 8.74$-$3 & 4.64$-$3 \\
   44   &  86 &   2.58$-$07 & 2.86$-$5 &  1.03$-$01  &  3.68$-$02  &  2.31$-$02  &  1.59$-$1  &  1.43$-$1  &  5.97$-$2   & 4.85$-$2  & 7.80$-$2 & 4.00$-$2 \\
   45   &  86 &   1.21$-$03 & 2.30$-$3 &  8.55$-$02  &  5.96$-$02  &  5.88$-$02  &  1.24$-$1  &  1.16$-$1  &  7.65$-$2   & 3.95$-$2  & 6.87$-$2 & 5.98$-$2 \\
   46   &  86 &   ........  &          &  1.43$-$01  &  4.19$-$02  &  2.03$-$02  &  2.07$-$1  &  1.85$-$1  &  7.10$-$2   & 6.90$-$2  & 1.08$-$1 & 4.63$-$2 \\
   47   &  86 &   1.89$-$05 & 5.52$-$3 &  1.68$-$01  &  1.39$-$01  &  1.46$-$01  &  3.08$-$1  &  2.87$-$1  &  1.86$-$1   & 3.86$-$2  & 7.68$-$2 & 9.50$-$2 \\
   48   &  86 &   1.77$-$01 & 1.73$-$1 &  4.10$+$00  &  5.97$+$00  &  7.13$+$00  &  3.71$-$0  &  4.12$-$0  &  6.24$-$0   & 1.73$-$0  & 3.82$-$0 & 6.65$-$0 \\
   49   &  86 &   ........  &          &  2.33$-$02  &  1.81$-$02  &  1.36$-$02  &  1.80$-$2  &  1.83$-$2  &  1.45$-$2   & 1.03$-$2  & 1.92$-$2 & 1.49$-$2 \\
   50   &  86 &   ........  &          &  3.98$-$02  &  3.08$-$02  &  2.31$-$02  &  3.05$-$2  &  3.10$-$2  &  2.46$-$2   & 1.77$-$2  & 3.30$-$2 & 2.54$-$2 \\
   51   &  86 &   ........  &          & 5.73$-$02  &  4.43$-$02  &  3.34$-$02   &  4.39$-$2  &  4.46$-$2  &  3.54$-$2   & 2.60$-$2  & 4.82$-$2  & 3.68$-$2 \\
   52   &  86 &   1.55$-$06 & 1.25$-$5 & 5.67$-$02  &  2.29$-$02  &  1.27$-$02   &  1.29$-$1  &  1.15$-$1  &  4.39$-$2   & 3.82$-$2  & 5.93$-$2  & 2.64$-$2 \\
   53   &  86 &   9.59$-$04 & 1.63$-$3 & 5.91$-$02  &  5.39$-$02  &  5.66$-$02   &  9.81$-$2  &  9.29$-$2  &  6.79$-$2   & 2.98$-$2  & 5.17$-$2  & 4.75$-$2 \\
   54   &  86 &   2.10$-$06 & 2.67$-$6 & 9.36$-$03  &  3.37$-$03  &  1.60$-$03   &  2.51$-$2  &  2.22$-$2  &  7.86$-$3   & 6.83$-$3  & 1.06$-$2  & 4.74$-$3 \\
   55   &  86 &   4.70$-$03 & 3.74$-$4 & 5.26$-$01  &  5.05$-$01  &  5.82$-$01   &  4.83$-$1  &  4.94$-$1  &  5.54$-$1   & 2.03$-$1  & 4.29$-$1  & 6.42$-$1 \\
   56   &  86 &   ........  &          & 1.67$-$02  &  2.07$-$03  &  1.05$-$03   &  1.63$-$2  &  1.43$-$2  &  5.08$-$3   & 3.86$-$3  & 5.85$-$3  & 2.48$-$3 \\
   57   &  86 &   ........  &          & 2.20$-$02  &  2.76$-$03  &  1.32$-$03   &  2.25$-$2  &  1.98$-$2  &  6.92$-$3   & 5.33$-$3  & 7.95$-$3  & 3.06$-$3 \\
   58   &  86 &   ........  &          & 2.03$-$02  &  4.33$-$03  &  3.44$-$03   &  7.17$-$3  &  6.84$-$3  &  4.33$-$3   & 2.76$-$3  & 4.79$-$3  & 3.28$-$3 \\
   59   &  86 &   ........  &          & 2.92$-$02  &  3.41$-$03  &  1.59$-$03   &  2.82$-$2  &  2.49$-$2  &  8.66$-$3   & 6.85$-$3  & 1.02$-$2  & 3.82$-$3 \\
   60   &  86 &   2.60$-$01 & 3.78$-$1 & 6.31$+$00  &  8.83$+$00  &  1.03$+$01   &  5.56$-$0  &  6.15$-$0  &  9.07$-$0   & 2.84$-$0  & 6.30$-$0  & 1.05$+$1 \\   
\hline
\end{tabular} 
}

\newpage
\clearpage
\begin{flushleft}
Table 6. Oscillator strengths (f, dimensionless), collision strengths ($\Omega$) and effective collision strengths ($\Upsilon$) for transitions of  Mg V with upper level J=86. $a{\pm}b \equiv$ $a\times$10$^{{\pm}b}$.
\end{flushleft}
{\tiny
\begin{tabular}{rrllllllllllll}                                                                                   
\hline   
& & Present f &  {\bf f of TS15 } & \multicolumn{3}{c}{Present $\Omega$} &  \multicolumn{3}{c}{Present $\Upsilon$}  &  \multicolumn{3}{c}{$\Upsilon$ of TS15}     \\                                                                                                                                                                                                   
\multicolumn{2}{c}{Transition} & &  & \multicolumn{3}{c}{Energy (Ryd)} & \multicolumn{3}{c}{Temperature (log, K)} & \multicolumn{3}{c}{Temperature (log, K)}\\                                           
\hline                                                                                                        
  $i$ & $j$ &       &      & 9.0         &    12.0     &    15.0     &   4.0       &  5.0        &  6.0         &   4.0        &  5.0        &  6.0       \\                                                                
\hline           
   61   &  86 &   ........  &          & 2.34$-$02  &  2.06$-$03  &  9.20$-$04   &  1.41$-$2  &  1.25$-$2  &  4.46$-$3   & 3.59$-$3  & 5.45$-$3  & 2.13$-$3 \\
   62   &  86 &   ........  &          & 2.86$-$02  &  2.70$-$03  &  1.21$-$03   &  1.72$-$2  &  1.53$-$2  &  5.53$-$3   & 4.63$-$3  & 7.06$-$3  & 2.81$-$3 \\
   63   &  86 &   ........  &          & 3.48$-$02  &  3.25$-$03  &  1.42$-$03   &  2.09$-$2  &  1.85$-$2  &  6.67$-$3   & 5.90$-$3  & 8.88$-$3  & 3.35$-$3 \\
   64   &  86 &   ........  &          & 3.38$-$02  &  7.13$-$03  &  6.66$-$03   &  3.26$-$2  &  2.92$-$2  &  1.33$-$2   & 4.23$-$3  & 7.25$-$3  & 5.82$-$3 \\
   65   &  86 &   ........  &          & 2.00$-$02  &  1.56$-$02  &  1.25$-$02   &  1.63$-$2  &  1.64$-$2  &  1.32$-$2   & 7.45$-$3  & 1.45$-$2  & 1.25$-$2 \\
   66   &  86 &   ........  &          & 4.03$-$02  &  3.10$-$02  &  2.41$-$02   &  2.96$-$2  &  3.03$-$2  &  2.48$-$2   & 1.57$-$2  & 3.07$-$2  & 2.58$-$2 \\
   67   &  86 &   ........  &          & 3.11$-$02  &  2.37$-$02  &  1.85$-$02   &  2.30$-$2  &  2.35$-$2  &  1.91$-$2   & 1.17$-$2  & 2.29$-$2  & 1.94$-$2 \\
   68   &  86 &   ........  &          & 1.49$-$01  &  9.89$-$02  &  9.59$-$02   &  1.28$-$1  &  1.26$-$1  &  1.04$-$1   & 4.59$-$2  & 8.92$-$2  & 9.30$-$2 \\
   69   &  86 &   ........  &          & 1.77$-$01  &  5.76$-$02  &  5.25$-$02   &  1.37$-$1  &  1.27$-$1  &  7.47$-$2   & 3.18$-$2  & 5.67$-$2  & 5.31$-$2 \\
   70   &  86 &   ........  &          & 9.02$-$02  &  4.36$-$02  &  3.27$-$02   &  8.41$-$2  &  7.92$-$2  &  4.61$-$2   & 2.90$-$2  & 5.19$-$2  & 3.47$-$2 \\
   71   &  86 &   ........  &          & 4.68$-$02  &  2.82$-$02  &  2.02$-$02   &  4.13$-$2  &  3.99$-$2  &  2.56$-$2   & 1.79$-$2  & 3.23$-$2  & 2.24$-$2 \\
   72   &  86 &   ........  &          & 1.55$-$02  &  9.49$-$03  &  6.82$-$03   &  1.31$-$2  &  1.27$-$2  &  8.38$-$3   & 5.97$-$3  & 1.09$-$2  & 7.69$-$3 \\
   73   &  86 &   ........  &          & 4.94$-$02  &  3.24$-$02  &  2.35$-$02   &  3.93$-$2  &  3.89$-$2  &  2.73$-$2   & 2.03$-$2  & 3.71$-$2  & 2.68$-$2 \\
   74   &  86 &   ........  &          & 3.54$-$01  &  2.46$-$01  &  2.66$-$01   &  3.13$-$1  &  3.07$-$1  &  2.78$-$1   & 1.05$-$1  & 2.09$-$1  & 2.57$-$1 \\
   75   &  86 &   ........  &          & 2.57$-$01  &  1.39$-$01  &  8.43$-$02   &  4.19$-$1  &  3.83$-$1  &  1.75$-$1   & 1.99$-$1  & 3.09$-$1  & 1.47$-$1 \\
   76   &  86 &   ........  &          & 2.01$-$01  &  1.09$-$01  &  6.67$-$02   &  3.26$-$1  &  2.98$-$1  &  1.37$-$1   & 1.55$-$1  & 2.41$-$1  & 1.15$-$1 \\
   77   &  86 &   ........  &          & 1.49$-$01  &  8.15$-$02  &  5.13$-$02   &  2.36$-$1  &  2.16$-$1  &  1.01$-$1   & 1.15$-$1  & 1.79$-$1  & 9.03$-$2 \\
   78   &  86 &   ........  &          & 2.94$-$02  &  1.85$-$02  &  1.21$-$02   &  6.50$-$2  &  5.89$-$2  &  2.61$-$2   & 2.84$-$2  & 4.34$-$2  & 2.05$-$2 \\
   79   &  86 &   ........  &          & 8.83$-$02  &  5.64$-$02  &  3.74$-$02   &  1.95$-$1  &  1.77$-$1  &  7.92$-$2   & 8.53$-$2  & 1.30$-$1  & 6.21$-$2 \\
   80   &  86 &   ........  &          & 1.44$-$01  &  9.11$-$02  &  6.03$-$02   &  3.20$-$1  &  2.90$-$1  &  1.29$-$1   & 1.47$-$1  & 2.26$-$1  & 1.15$-$1 \\
   81   &  86 &   ........  &          & 2.39$-$01  &  1.56$-$01  &  1.41$-$01   &  3.04$-$1  &  2.86$-$1  &  1.83$-$1   & 1.72$-$1  & 3.03$-$1  & 2.68$-$1 \\
   82   &  86 &   ........  &          & 4.76$-$02  &  2.10$-$02  &  1.09$-$02   &  1.16$-$1  &  1.03$-$1  &  3.92$-$2   & 4.25$-$2  & 6.17$-$2  & 2.44$-$2 \\
   83   &  86 &   ........  &          & 1.11$-$01  &  5.02$-$02  &  2.72$-$02   &  2.78$-$1  &  2.47$-$1  &  9.45$-$2   & 1.01$-$1  & 1.46$-$1  & 5.71$-$2 \\
   84   &  86 &   ........  &          & 1.48$+$00  &  1.14$+$00  &  1.15$+$00   &  1.21$-$0  &  1.22$-$0  &  1.15$-$0   & 5.34$-$1  & 1.03$-$0  & 1.11$-$0 \\
   85   &  86 &   ........  &          & 2.86$-$01  &  1.99$-$01  &  2.06$-$01   &  3.20$-$1  &  3.07$-$1  &  2.33$-$1   & 1.16$-$1  & 2.09$-$1  & 2.27$-$1 \\
\hline  
\end{tabular}
}
\begin{flushleft}
{\small
TS15: Results of Tayal and Sossah \cite{sst}  for f-values with the MCHF and for $\Upsilon$  with the {\sc BSR} code \\ 
}
\end{flushleft}


\newpage
\clearpage
\begin{figure*}
\includegraphics[angle=-90,width=0.9\textwidth]{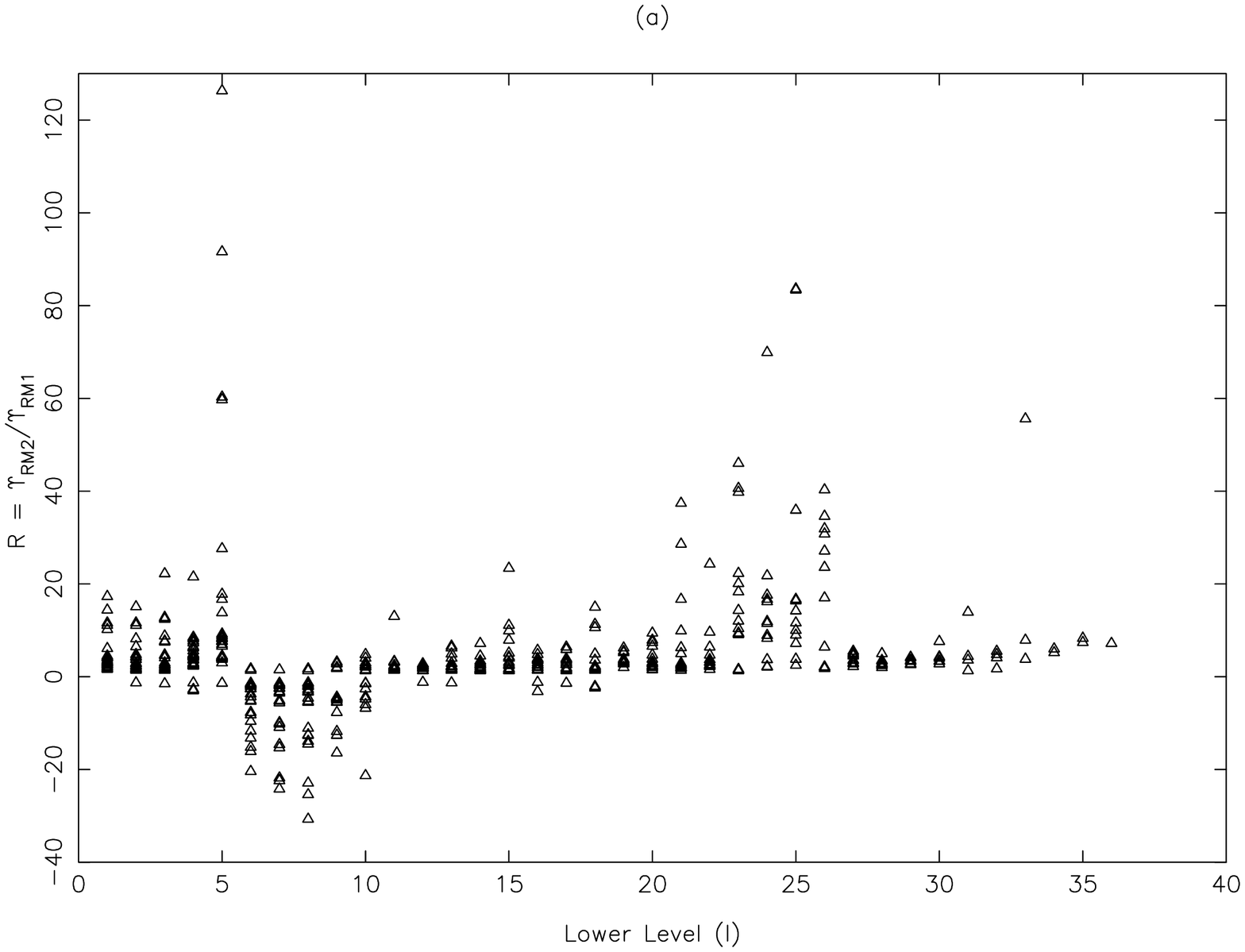}
 \vspace{-1.5cm}
 \caption{Comparison of RM1 (Hudson et al. \cite{hud}) and RM2  (Tayal and Sossah\cite{sst}) values of $\Upsilon$ for transitions of Mg~V at (a) T$_e$ = 10$^{4}$ K, (b) T$_e$ = 10$^{5}$ K,  and (c) T$_e$ = 10$^{6}$ K. Negative R values plot  $\Upsilon_{\rm RM1}$/$\Upsilon_{\rm RM2}$ and indicate that $\Upsilon_{\rm RM1}$ $>$ $\Upsilon_{\rm RM2}$. Only those transitions are shown which differ by over 20\%.}
 \end{figure*}

\setcounter{figure}{0}
 \begin{figure*}
\includegraphics[angle=-90,width=0.9\textwidth]{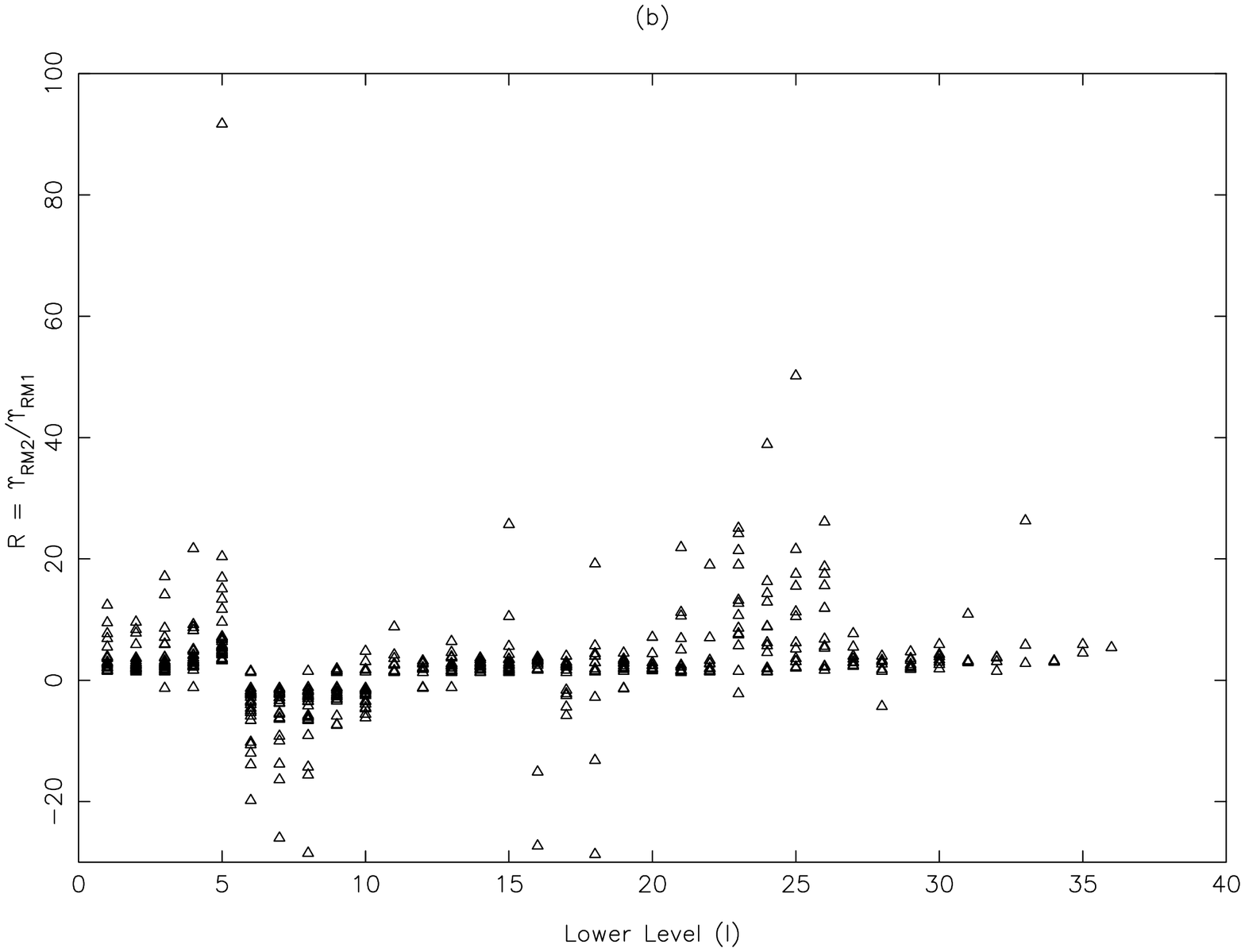}
 \vspace{-1.5cm}
\caption{continued.}
 \end{figure*}

\setcounter{figure}{0}
 \begin{figure*}
\includegraphics[angle=-90,width=0.9\textwidth]{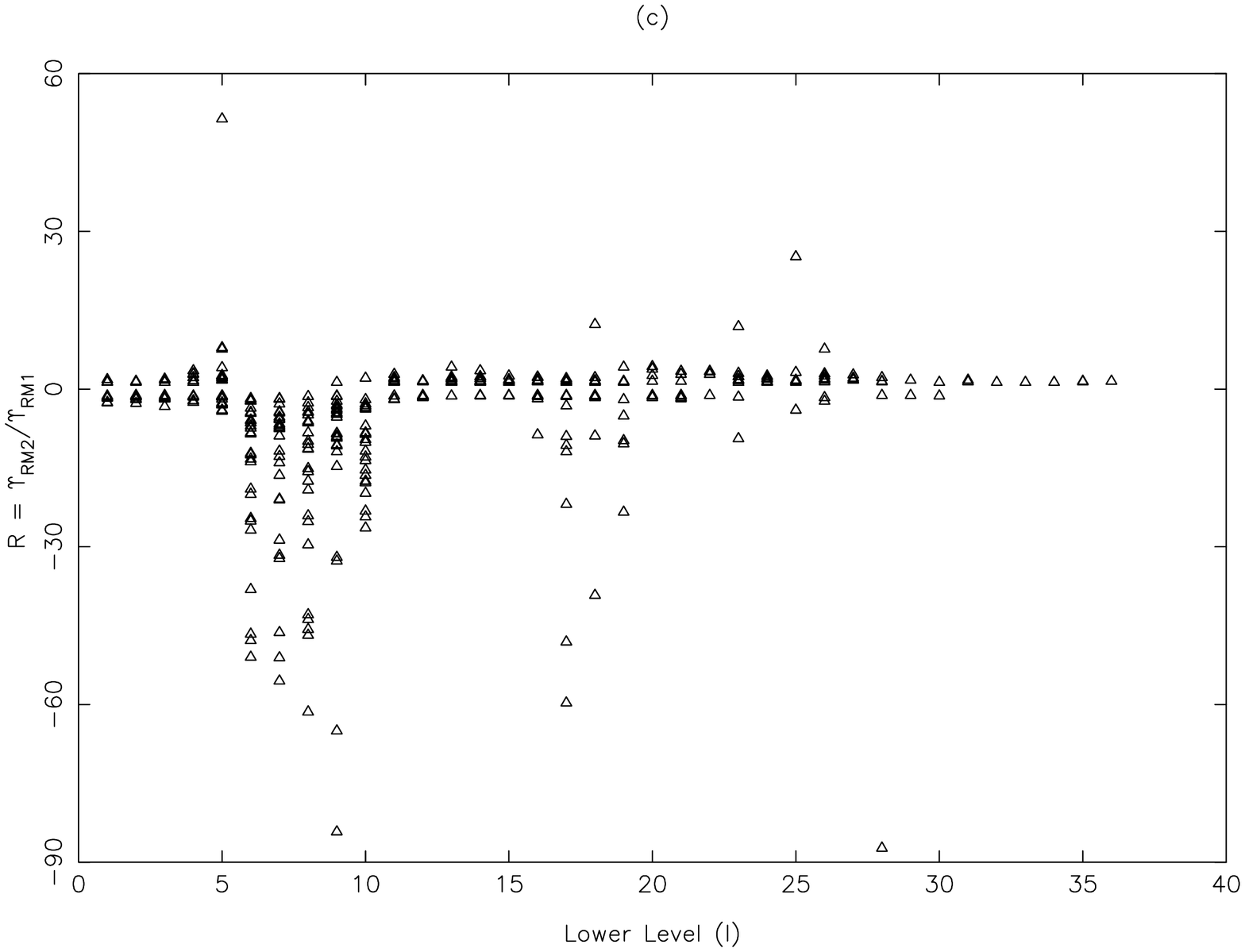}
 \vspace{-1.5cm}
 \caption{continued.}
 \end{figure*}

 \begin{figure*}
\includegraphics[angle=-90,width=0.9\textwidth]{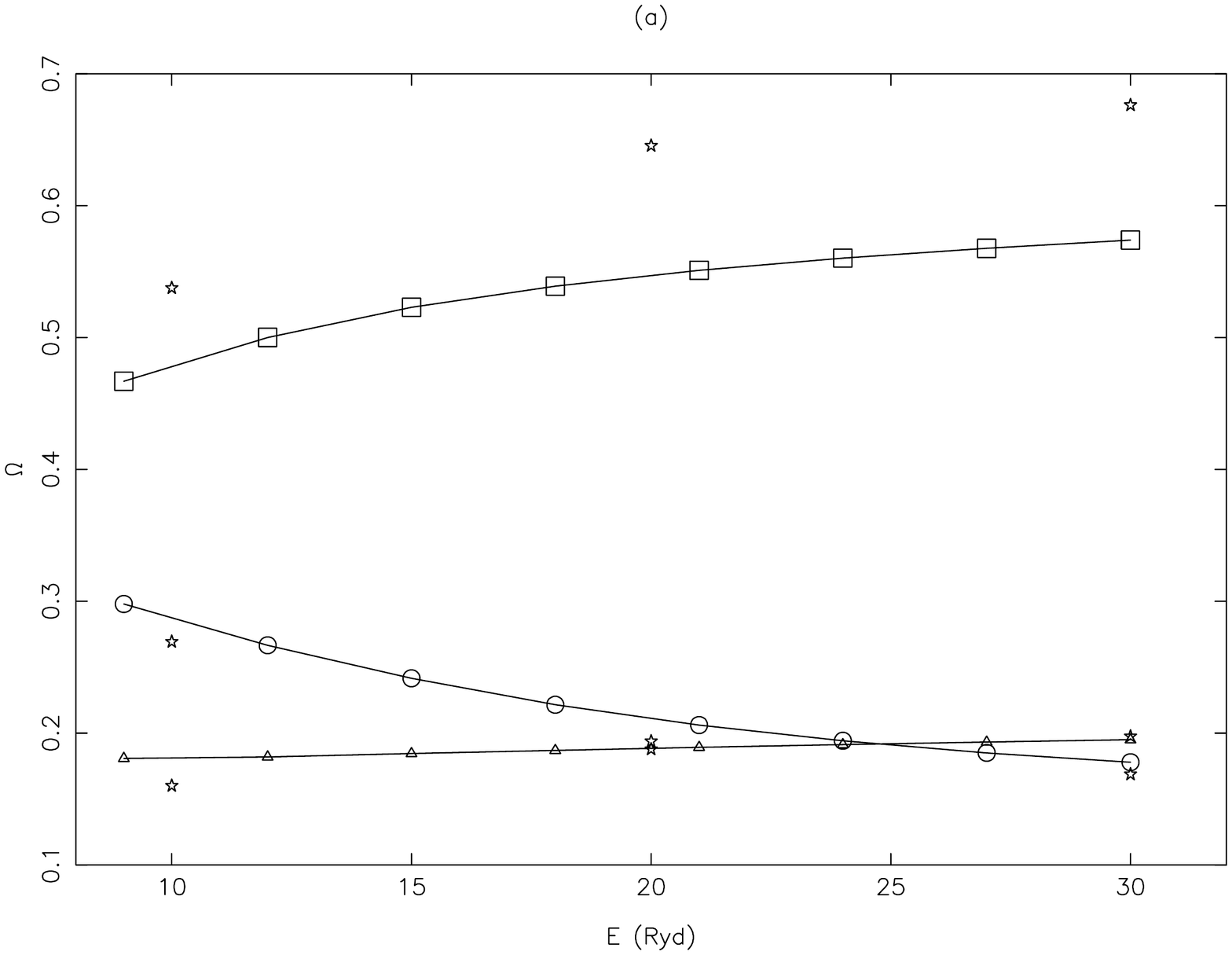}
 \vspace{-1.5cm}
 \caption{Comparison of collision strengths for the (a) 1--2 (circles: 2s$^2$2p$^4$~$^3$P$_2$ -- 2s$^2$2p$^4$~$^3$P$_1$), 1--22 (triangles: 2s$^2$2p$^4$~$^3$P$_2$ -- 2s$^2$2p$^3$($^4$S)3p~$^3$P$_2$) and 1--42 (squares: 2s$^2$2p$^4$~$^3$P$_2$ -- 2s$^2$2p$^3$($^2$D)3p~$^3$P$_2$) forbidden transitions of Mg~V. Continuous curves: present results with {\sc darc}, stars: DW results of  Bhatia et al. \cite{ble}.}
 \end{figure*}
 
\setcounter{figure}{1} 
 \begin{figure*}
\includegraphics[angle=-90,width=0.9\textwidth]{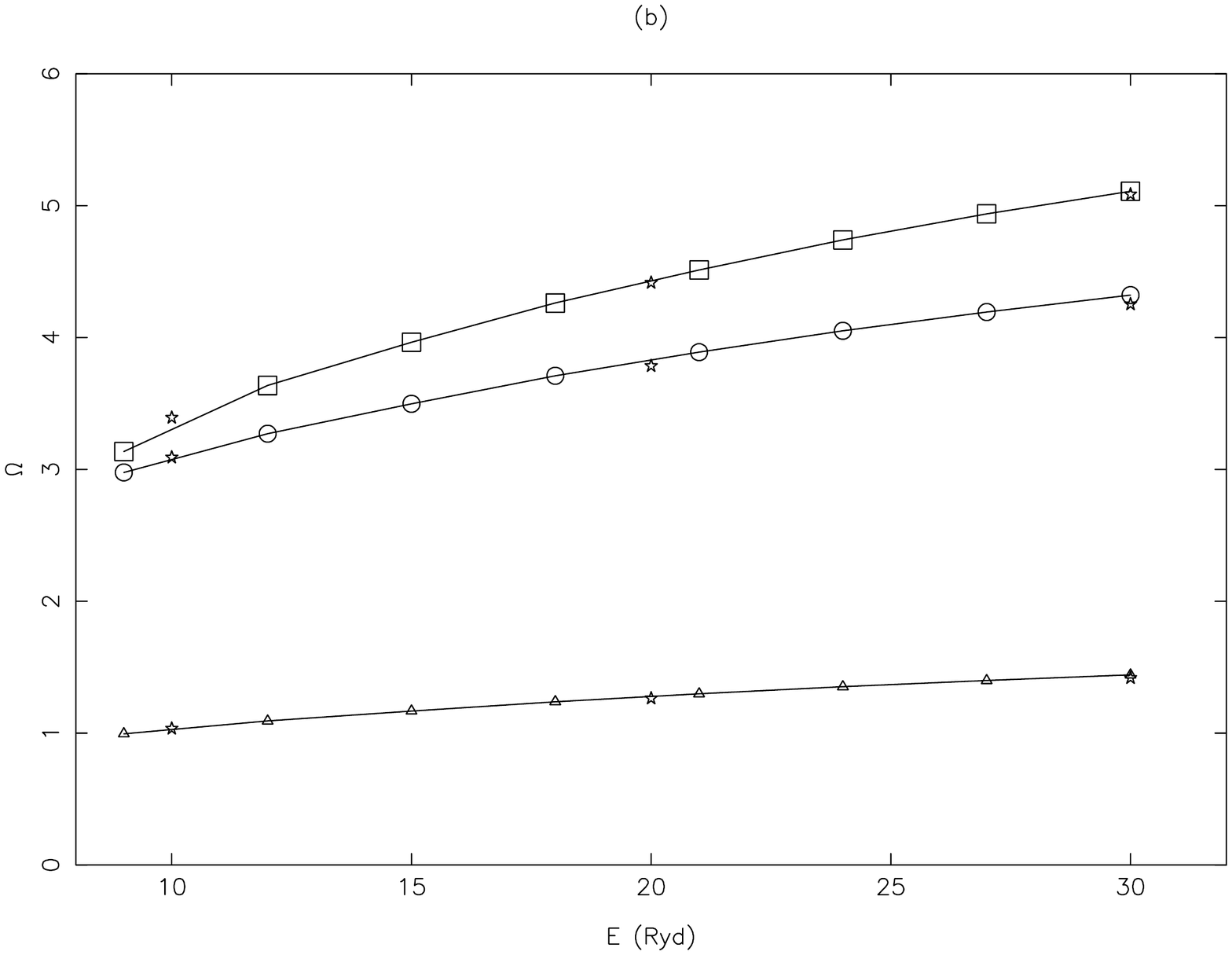}
 \vspace{-1.5cm}
\caption{Comparison of collision strengths for the (a) 1--2 (circles: 2s$^2$2p$^4$~$^3$P$_2$ -- 2s$^2$2p$^4$~$^3$P$_1$), 1--22 (triangles: 2s$^2$2p$^4$~$^3$P$_2$ -- 2s$^2$2p$^3$($^4$S)3p~$^3$P$_2$) and 1--42 (squares: 2s$^2$2p$^4$~$^3$P$_2$ -- 2s$^2$2p$^3$($^2$D)3p~$^3$P$_2$) forbidden,  and (b) 1--6 (circles: 2s$^2$2p$^4$~$^3$P$_2$ -- 2s2p$^5$~$^3$P$^o_2$), 1--7 (triangles: 2s$^2$2p$^4$~$^3$P$_2$ -- 2s2p$^5$~$^3$P$^o_1$) and 4--9 (squares: 2s$^2$2p$^4$~$^1$D$_2$ -- 2s2p$^5$~$^1$P$^o_1$) allowed transitions of Mg~V. Continuous curves: present results with {\sc darc}, stars: DW results of  Bhatia et al. \cite{ble}.}
 \end{figure*}

\begin{figure*}
\includegraphics[angle=90,width=0.9\textwidth]{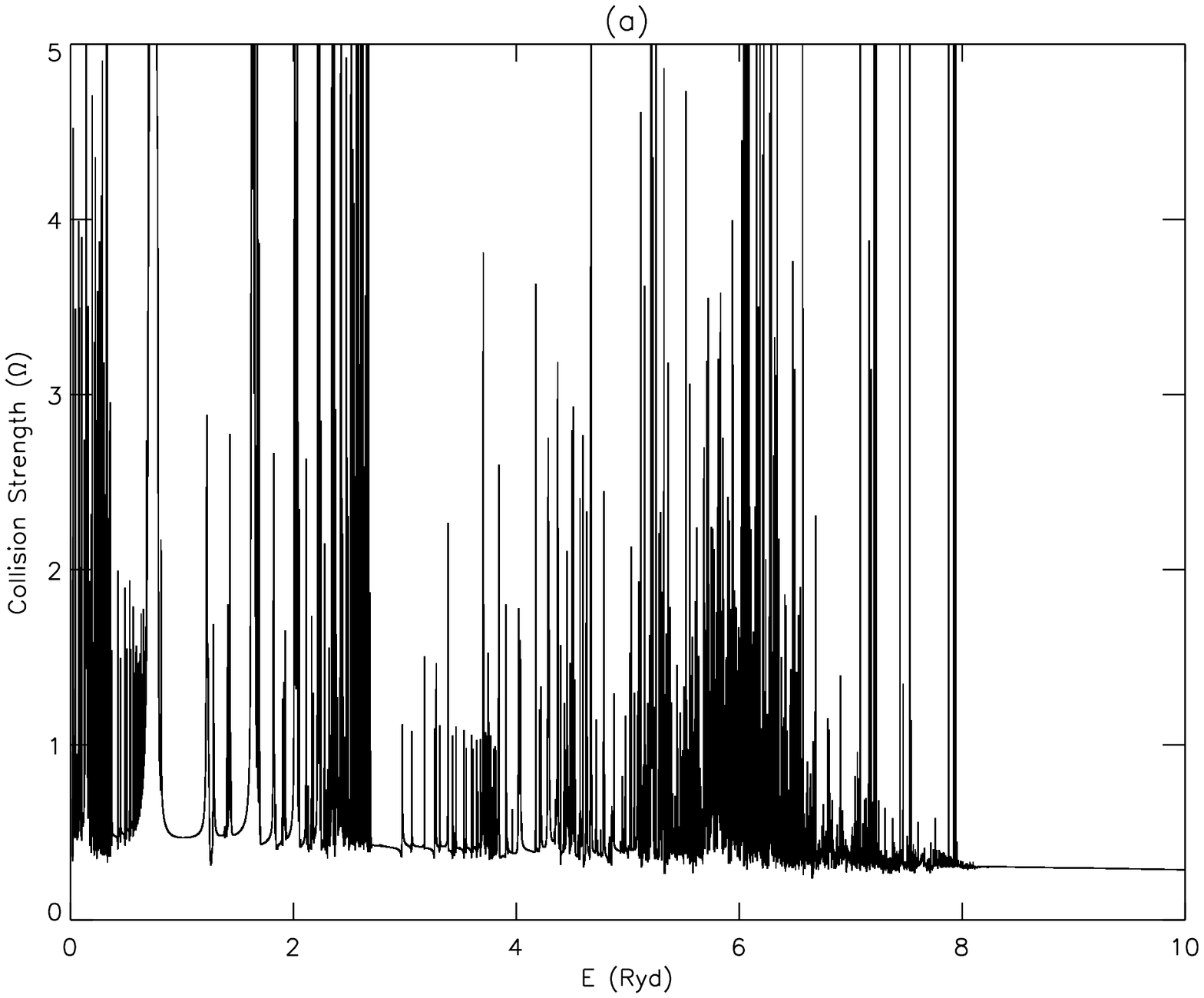}
 \vspace{-1.5cm}
 \caption{Collision strengths for the (a) 1--2 (2s$^2$2p$^4$~$^3$P$_2$ -- 2s$^2$2p$^4$~$^3$P$_1$), (b) 1--3 (2s$^2$2p$^4$~$^3$P$_2$ -- 2s$^2$2p$^4$~$^3$P$_0$), and (c)  4--9 (2s$^2$2p$^4$~$^1$D$_2$ -- 2s2p$^5$~$^1$P$^o_1$) transitions of Mg V.}
 \end{figure*}
 
 \setcounter{figure}{2}
 \begin{figure*}
\includegraphics[angle=90,width=0.9\textwidth]{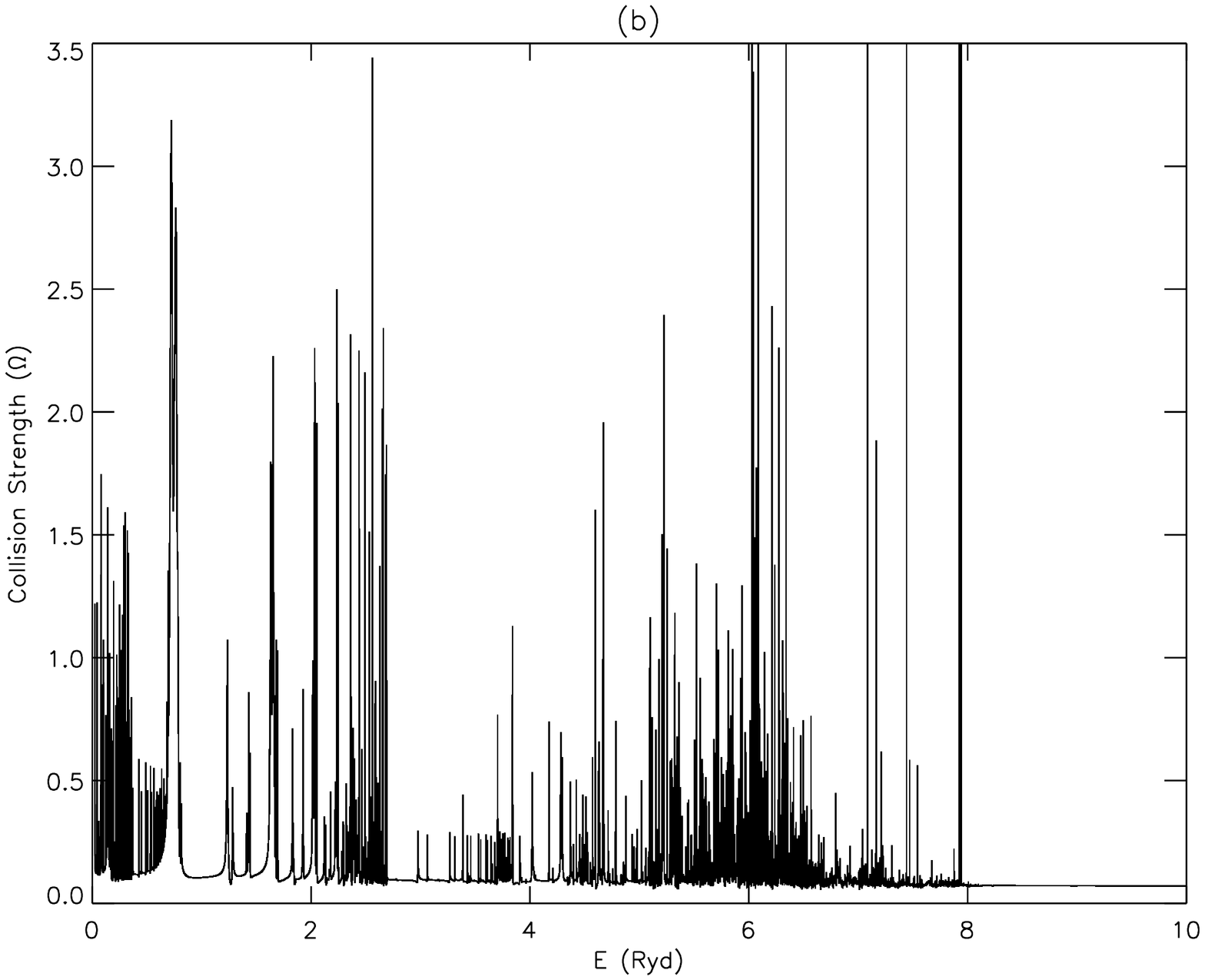}
 \vspace{-1.5cm}
 \caption{continued.}
 \end{figure*}
 
 \setcounter{figure}{2}
 \begin{figure*}
\includegraphics[angle=90,width=0.9\textwidth]{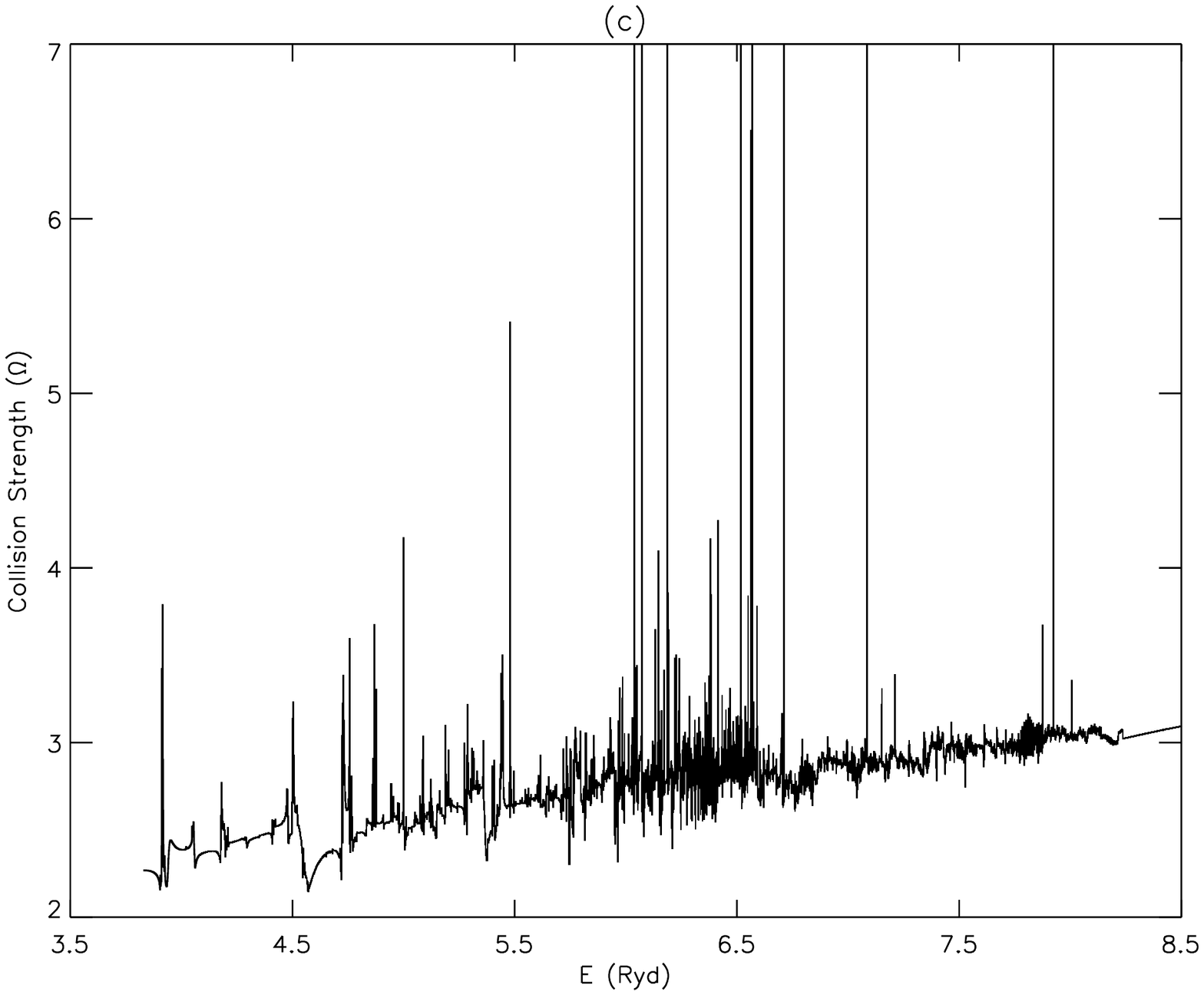}
 \vspace{-1.5cm}
 \caption{continued.}
 \end{figure*}

 \begin{figure*}
\includegraphics[angle=-90,width=0.9\textwidth]{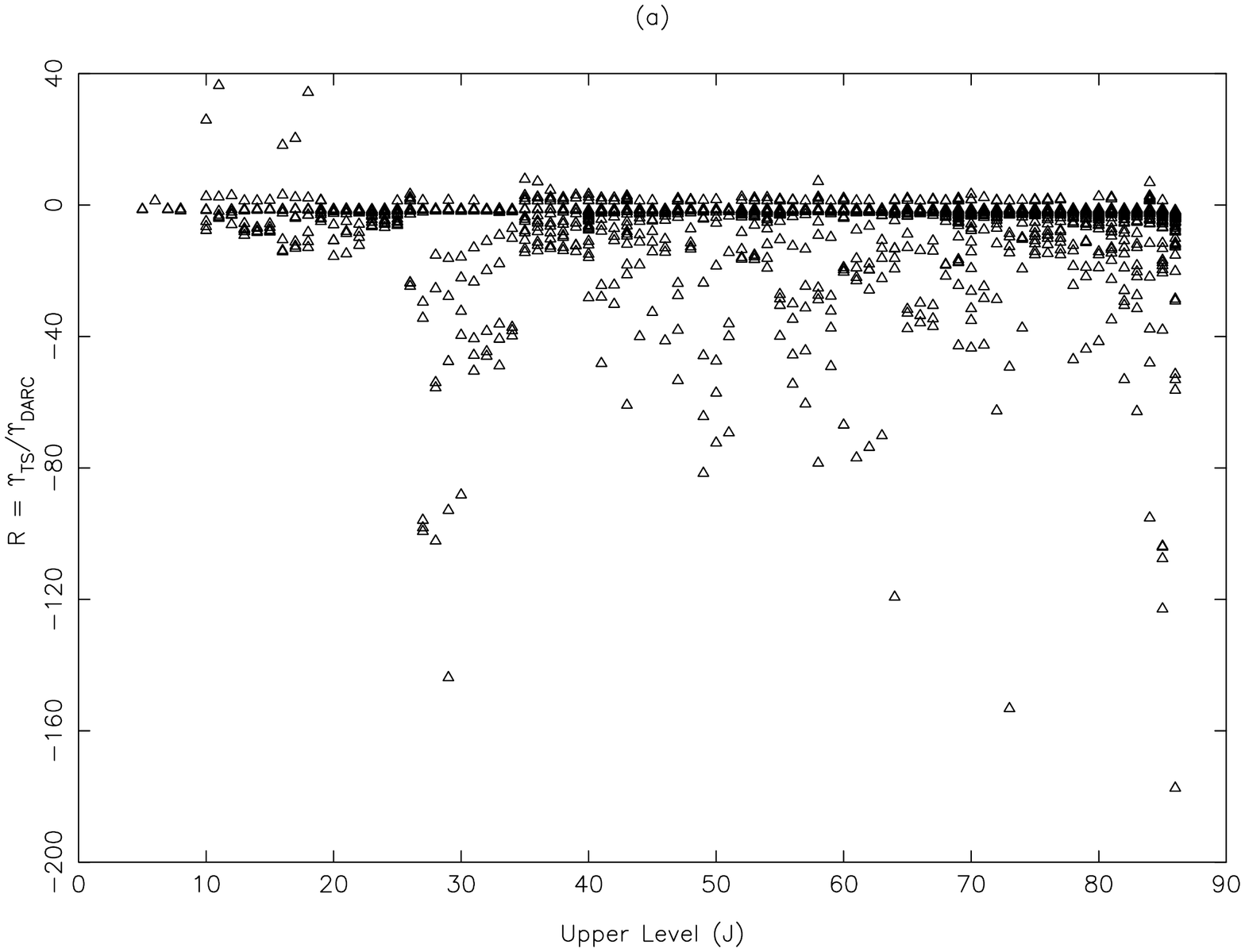}
 \vspace{-1.5cm}
\caption{Comparisons of  $\Upsilon$ between our results with {\sc darc} and those of Tayal and Sossah (\cite{sst}: TS) for transitions of Mg~V at (a) T$_e$ = 10$^4$~K, (b) T$_e$ = 10$^5$~K, and (c) T$_e$ = 10$^6$~K. Negative R values indicate that $\Upsilon_{\rm DARC}$ $>$ $\Upsilon_{\rm TS}$. Only those transitions are shown which differ by over 20\%.}
 \end{figure*}

\setcounter{figure}{3} 
 \begin{figure*}
\includegraphics[angle=-90,width=0.9\textwidth]{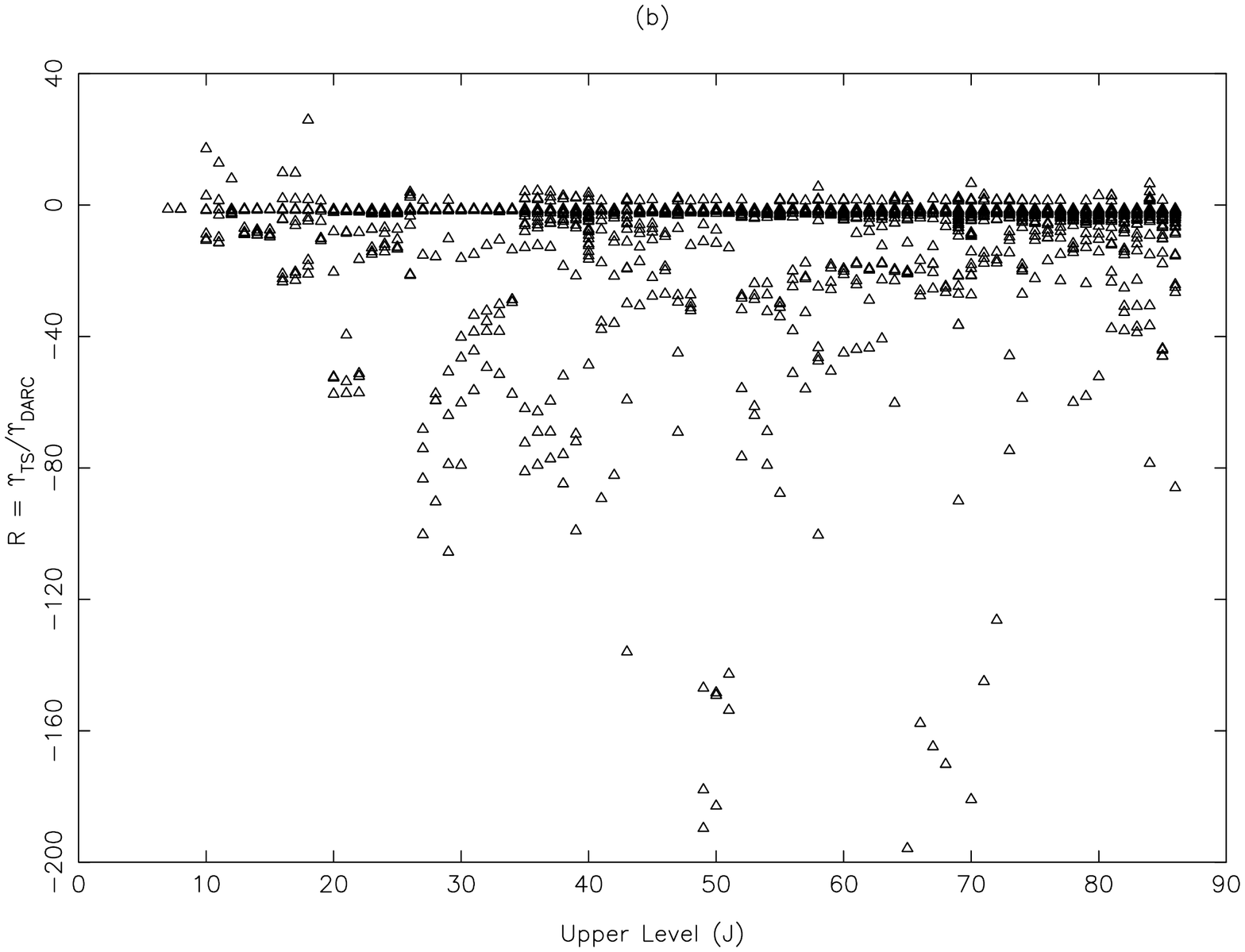}
 \vspace{-1.5cm}
 \caption{continued.}
 \end{figure*}

\setcounter{figure}{3} 
 \begin{figure*}
\includegraphics[angle=-90,width=0.9\textwidth]{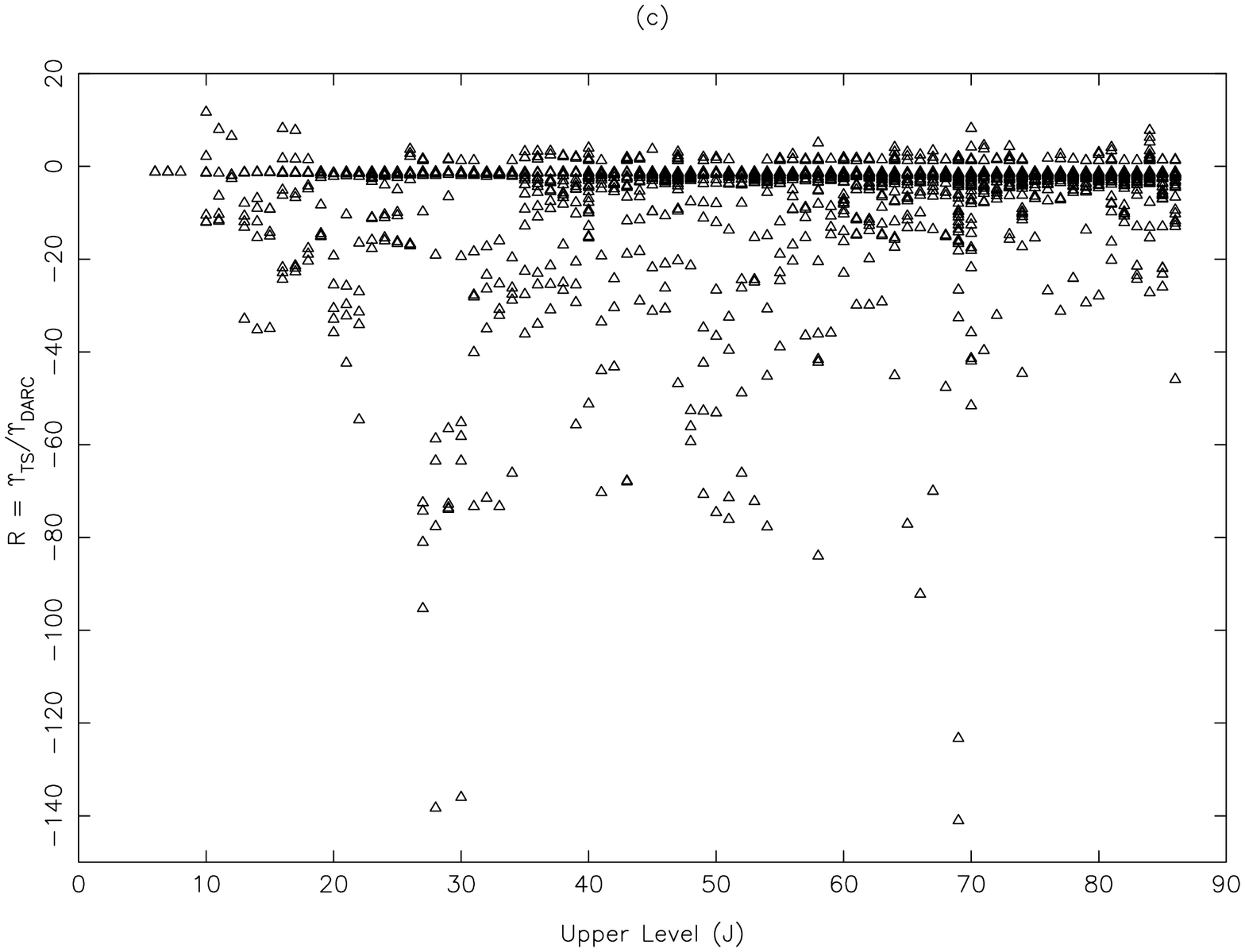}
 \vspace{-1.5cm}
 \caption{continued.}
 \end{figure*}

\begin{figure*}
\includegraphics[angle=-90,width=0.9\textwidth]{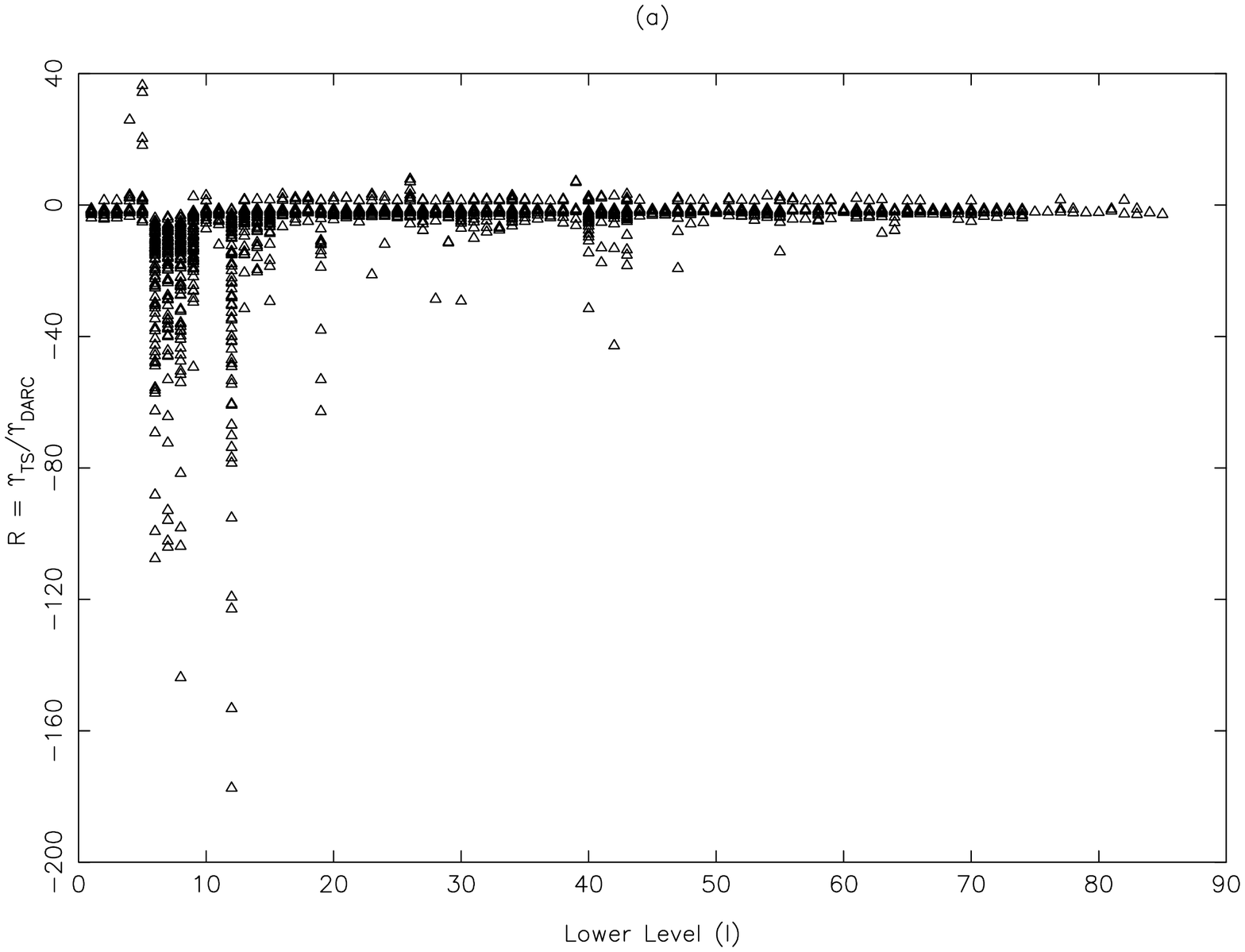}
 \vspace{-1.5cm}
 \caption{Comparisons of  $\Upsilon$ between our results with {\sc darc} and those of Tayal and Sossah (\cite{sst}: TS) for transitions of Mg~V at (a) T$_e$ = 10$^4$~K, (b) T$_e$ = 10$^5$~K and (c) T$_e$ = 10$^6$~K. Negative R values indicate that $\Upsilon_{\rm DARC}$ $>$ $\Upsilon_{\rm TS}$. Only those transitions are shown which differ by over 20\%.}
 \end{figure*}

\setcounter{figure}{4}
 \begin{figure*}
\includegraphics[angle=-90,width=0.9\textwidth]{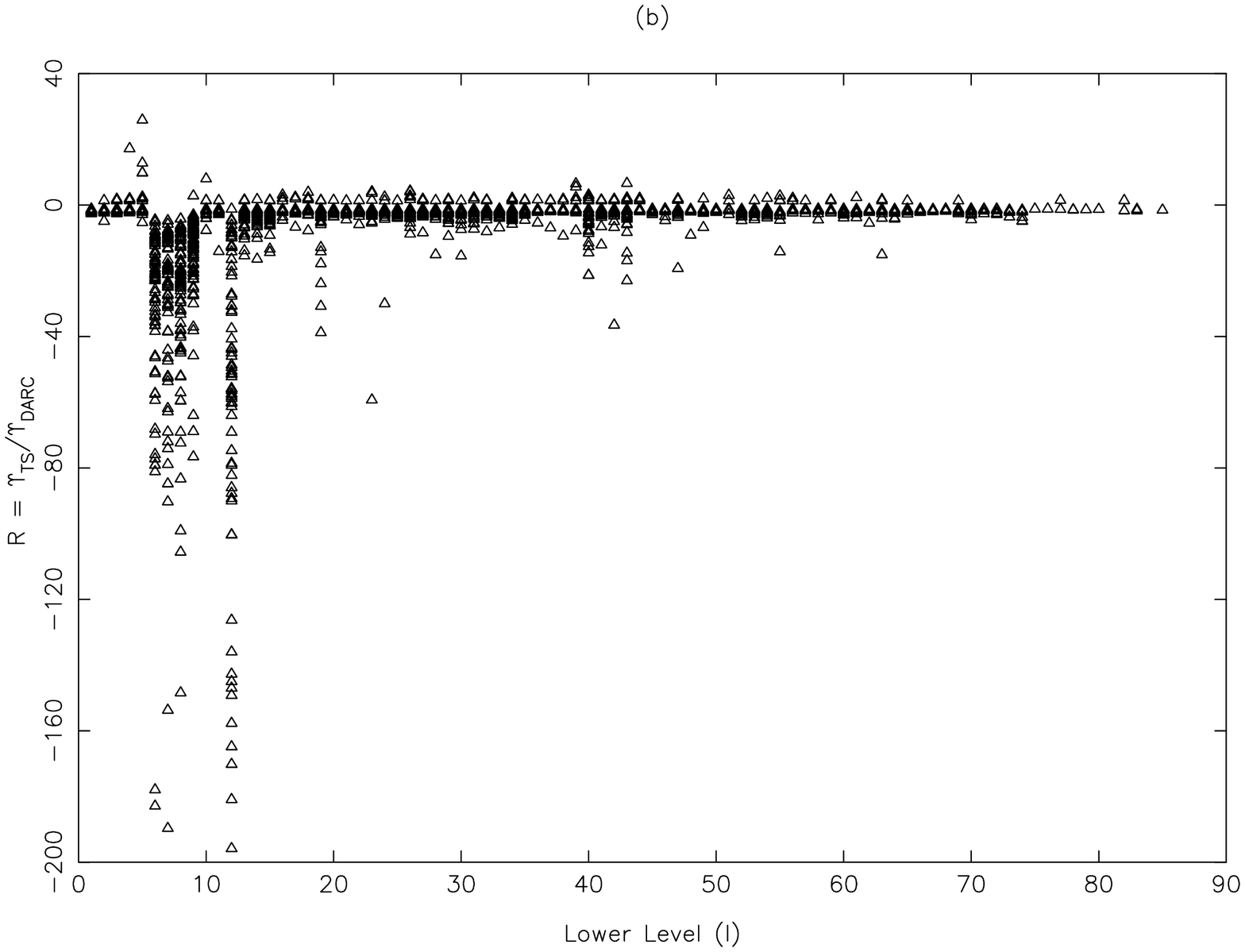}
 \vspace{-1.5cm}
\caption{continued.}
 \end{figure*}

\setcounter{figure}{4}
 \begin{figure*}
\includegraphics[angle=-90,width=0.9\textwidth]{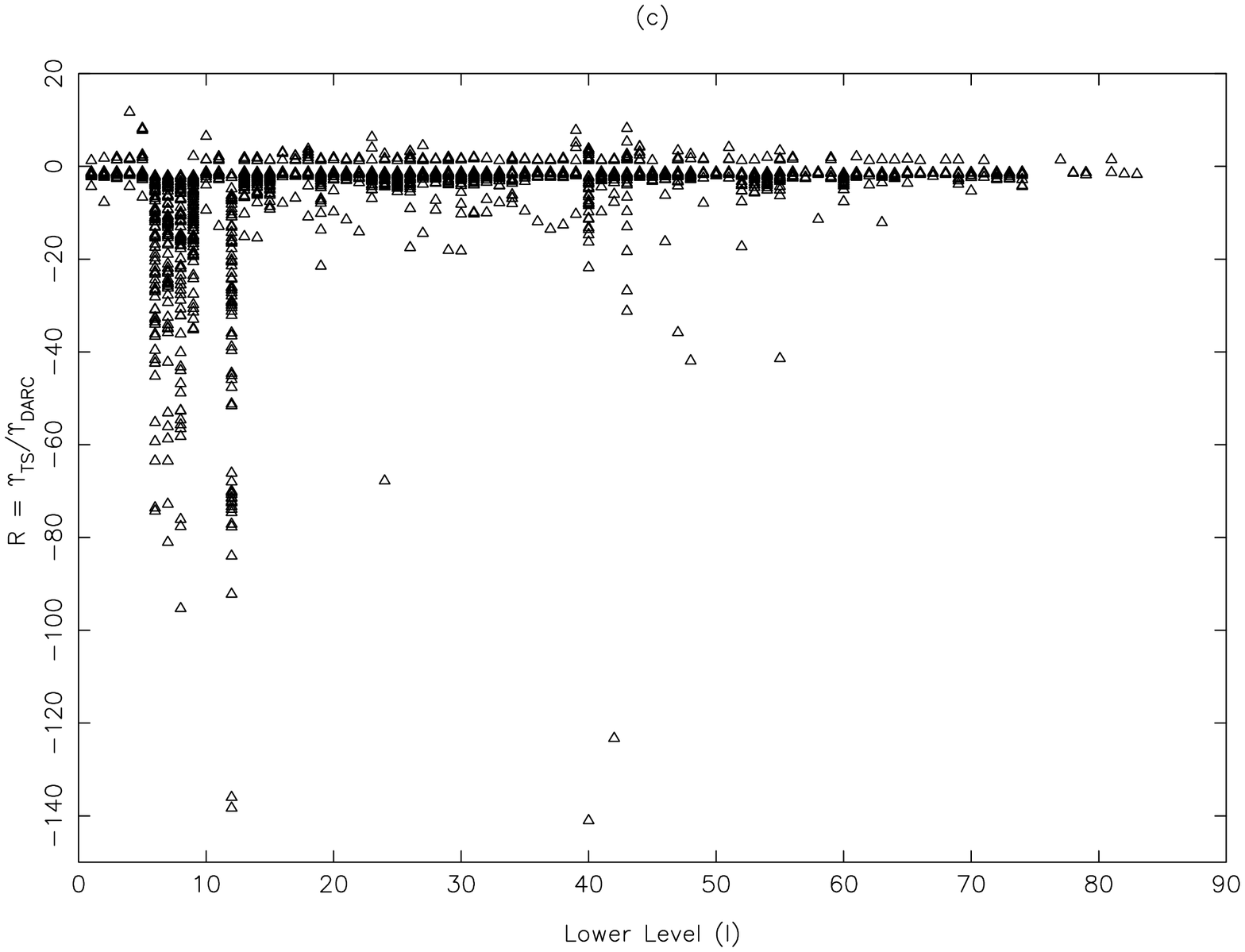}
 \vspace{-1.5cm}
 \caption{continued.}
 \end{figure*}

 \begin{figure*}
\includegraphics[angle=90,width=0.9\textwidth]{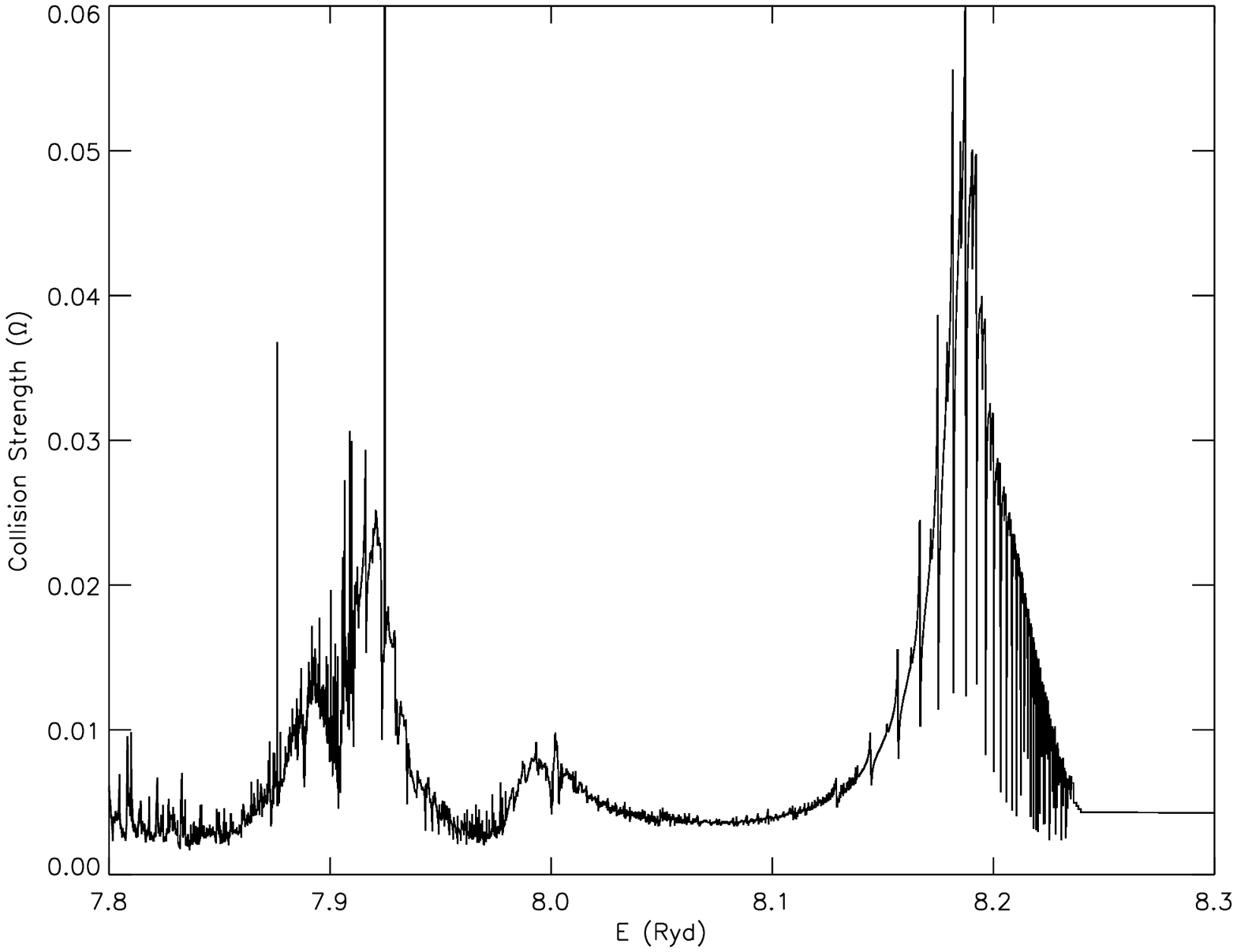}
 \vspace{-1.5cm}
 \caption{Collision strengths for the  12--30 (2p$^6$~$^1$S$_0$ -- 2p$^3$3p~$^3$D$_3$) transition of Mg V.}
 \end{figure*}

\end{document}